\documentclass[conference]{IEEEtran}
\usepackage{fancyhdr}
\IEEEoverridecommandlockouts
\usepackage{amsmath,amssymb,amsfonts}
\usepackage{algorithmic}
\usepackage{graphicx}
\usepackage{textcomp}
\usepackage{xcolor}
\usepackage{dcolumn}
\usepackage{subcaption}
\usepackage{caption}
\usepackage{booktabs}
\usepackage[version=4]{mhchem}
\usepackage{url}
\usepackage[noadjust]{cite} 
\usepackage{balance}

\newcolumntype{d}[1]{D{.}{.}{#1}}
\newcommand{\tabcenter}[1]{\multicolumn{1}{l}{#1}}

\def\BibTeX{{\rm B\kern-.05em{\sc i\kern-.025em b}\kern-.08em
    T\kern-.1667em\lower.7ex\hbox{E}\kern-.125emX}}
\begin{document}

\title{Pushing the limit of molecular dynamics with \textit{{ab initio}} accuracy  to 100 million atoms with machine learning
}

\author{
  \IEEEauthorblockN{
    Weile Jia\IEEEauthorrefmark{1},
    Han Wang\IEEEauthorrefmark{2},
    Mohan Chen\IEEEauthorrefmark{3},
    Denghui Lu\IEEEauthorrefmark{3},
    Lin Lin\IEEEauthorrefmark{1}\IEEEauthorrefmark{5},
    Roberto Car\IEEEauthorrefmark{6},
    Weinan E\IEEEauthorrefmark{6},
    Linfeng Zhang\IEEEauthorrefmark{6} \textsuperscript{\textsection}
  \IEEEauthorblockA{
    \IEEEauthorrefmark{1}University of California, Berkeley, Berkeley, USA \\
      Email: jiaweile@berkeley.edu, linlin@math.berkeley.edu}
    \IEEEauthorblockA{\IEEEauthorrefmark{2} Laboratory of Computational Physics, Institute of Applied Physics and Computational Mathematics, Beijing, China\\
      Email: wang\_han@iapcm.ac.cn}
    \IEEEauthorblockA{\IEEEauthorrefmark{3}CAPT, HEDPS,College of Engineering, Peking University,
Beijing, China
 \\Email: mohanchen@pku.edu.cn, denghuilu@pku.edu.cn}
    \IEEEauthorblockA{\IEEEauthorrefmark{5}Lawrence Berkeley National Laboratory, Berkeley, USA }
     \IEEEauthorblockA{\IEEEauthorrefmark{6} Princeton University, Princeton, USA \\
 Email: rcar@princeton.edu, weinan@math.princeton.edu,linfengz@princeton.edu}
 }
}
\maketitle
\begingroup\renewcommand\thefootnote{\textsection}
\footnotetext{Corresponding author}
\endgroup
\thispagestyle{fancy}
\lhead{}
\rhead{}
\chead{}
\lfoot{\footnotesize{
SC20, November 9-19, 2020, Is Everywhere We Are
\newline 978-1-7281-9998-6/20/\$31.00 \copyright 2020 IEEE}}
\rfoot{}
\cfoot{}
\renewcommand{\headrulewidth}{0pt}
\renewcommand{\footrulewidth}{0pt}

\begin{abstract}
%\WH{1, How many folds of accelerations have been achieved. Comparing to the original code. 2, Key techniques used in our work. 3, Peak performance. 4, What does it mean to applications. } \WL{place holder, please revise based on this}
%\LZ{Notice: 150-word limit. Right now we have 155 words. I add some suggested changes for your consideration.}
For 35 years, \textit{ab initio} molecular dynamics (AIMD) has been the method of choice for modeling complex atomistic phenomena from first principles. 
However, most AIMD applications are limited by computational cost to systems with thousands of atoms at most. 
We report that a machine learning-based simulation protocol (Deep Potential Molecular Dynamics), while retaining \textit{ab initio} accuracy, can
simulate more than 1 nanosecond-long trajectory of over 100 million atoms per day, using
a highly optimized code (GPU DeePMD-kit) on the Summit supercomputer. 
Our code can efficiently scale up to the entire Summit supercomputer, attaining 91 PFLOPS in double precision (45.5\% of the peak) and {162/275 PFLOPS in mixed-single/half precision}. 
The great accomplishment of this work is that it opens the door to simulating unprecedented size and time scales with \textit{ab initio} accuracy. 
It also poses new challenges to the next-generation supercomputer for a better integration of machine learning and physical modeling.% (147 words.)
\end{abstract}


\section{Justification for Prize}
%\WL{GB requirement for this section: 50 words max}
%old version
%Record molecular dynamics simulation of >100 million atoms with \textit{ab initio} accuracy. For a 127-million-atom copper system, time-to-solution of $7.3\times 10^{-10}$ s/step/atom, or equivalently one nanosecond/day,  >$1000\times$ improvement w.r.t state-of-the-art. Double precision performance of 91 PFLOPS on 4,560 nodes of Summit ($45.5\%$ of the peak); mixed precision of 162 PFLOPS.
%Record molecular dynamics simulation of >100 million atoms with \textit{ab initio} accuracy. For a 127-million-atom copper system, time-to-solution of $2.68\times 10^{-10}$ s/step/atom, or equivalently $2.5$ nanosecond/day,  >$1000\times$ improvement w.r.t state-of-the-art. Double precision performance of 91 PFLOPS on Summit ($45.5\%$ of the peak); 162/275 PFLOPS for mixed single/half precision.\WL{Note:Time-to-solution is calculated using MIX-16 version.For double, it is 29 hours per nanosecond, $8.1\times10^{-10}$}
Record molecular dynamics simulation of $>$100 million atoms with \textit{ab initio} accuracy. Double/mixed-single/mixed-half precision performance of 91/162/275 PFLOPS on 4,560 nodes of Summit (27,360 GPUs). For a 127-million-atom copper system, time-to-solution of 8.1/4.6/2.7$\times10^{-10}$ s/step/atom, or equivalently $0.8/1.5/2.5$ nanosecond/day,  $>$1000$\times$ improvement w.r.t state-of-the-art.

\section{Performance Attributes} \label{sec:algorithm}
%\WL{GB requirement for this section: use a table listing each attribute title and value in a separate row) Category of achievement (1+ of:  scalability, time-to-solution, peak performance) Type of method used (1 of:  explicit, implicit, both explicit and implicit, semi-implicit, n/a)   Results reported on the basis of (1 of:  whole application including I/O; whole application except I/O; kernel only; other [specify]) Precision reported (1 of:  single precision, double precision, mixed precision)  System scale (1 of:  results measured on full-scale system, projected from results of smaller system, other [specify])  Measurement mechanism (1 of:  timers, FLOP count, static analysis tool, performance modeling, other [specify] }
%\WL{might need double confirm the "category of achievements" and "type of method used."}

{\footnotesize
\begin{center}
\begin{tabular}{l l} 
 \toprule
 Performance attribute & Our submission  \\
 \midrule
 Category of achievement  & Time-to-solution, scalability \\ 
 Type of method used  &  Deep potential molecular dynamics \\
 Results reported on basis of & Whole application including I/O \\
 Precision reported & Double precision, mixed precision \\
 System scale & Measured on full system \\
 Measurements  & Timers, FLOP count\\
\bottomrule
\end{tabular}
\end{center}
}
\vskip 1em

\section{Overview of the Problem}
%\WL{ GB requirement for this section: description of the problem and its importance, in terms understandable to a non-specialist (1 p max)  }
\subsection{\textit{ab initio} molecular dynamics}
Molecular dynamics (MD)~\cite{frenkel2001understanding,tuckerman2010statistical} is an \textit{in silico} simulation tool
for describing atomic processes that occur in materials and molecules.
%It is based on numerically integrating the trajectories of the atoms according to Newton's laws of motion.
The accuracy of MD lies in the description of atomic interactions, for which the \textit{ab initio} molecular dynamics (AIMD) scheme~\cite{car1985unified,marx2009ab} stands out by evolving atomic systems with the interatomic forces generated on-the-fly using first-principles electronic structure methods such as the density functional theory (DFT)~\cite{64PR-Hohenberg,kohn1965self}. 
AIMD permits chemical bond cleavage and formation events to occur and accounts for electronic polarization effects.
Due to the faithful description of atomic interactions by DFT, AIMD has been the major avenue for the microscopic understanding of 
%the macroscopic phenomena in various disciplines such as physics, chemistry, and material science.
%AIMD has been successfully applied to 
a broad spectrum of issues, such as %biophysics and life science~\cite{carloni2002role}, 
drug discovery~\cite{carloni2002role,aminpour2019overview}, complex chemical processes~\cite{leung2005ab,chen2018hydroxide}, nanotechnology~\cite{raty2005growth}, etc.

The computational cost of AIMD {generally} scales cubically with respect to the number of electronic degrees of freedom.
%, or roughly speaking the number of atoms in the system. 
On a desktop workstation, the typical spatial and temporal scales achievable by AIMD are $\sim$100 atoms and $\sim$10 picoseconds. %($10^{-11}$ second). 
%The largest system handled by modern high performance computers (HPCs) like Summit, using a typical cubic-scaling algorithm, is composed of 10,508 Magnesium atoms (105K valence electrons)~\cite{das2019fast}.
%In the past 14 years (2006 to 2019), 
From 2006 to 2019,
the peak performance of the world's fastest supercomputer has increased about 550-folds, (from 360 TFLOPS of BlueGene/L to 200 PFLOPS of Summit), but the accessible system size has only increased 8 times (from 1K Molybdenum atoms with 12K valence electrons
~\cite{gygi2006large} to 11K Magnesium atoms with 105K valence electrons~\cite{das2019fast}), which obeys almost perfectly the cubic-scaling law. 
Linear-scaling DFT methods~\cite{wang2008linearly,eisenbach2009scalable,rozanski2016linear,nakata2020large}
%, which scale linearly with the system size, 
have been under active developments, yet the pre-factor in the complexity is still large, and 
%the codes are not suitable for long-time MD simulation.
the time scales attainable in MD simulations remain rather short.

For problems in complex chemical reactions~\cite{nakano2007divide,li2015revealing}, electrochemical cells~\cite{jorn2013atomistic}, nanocrystalline materials~\cite{98N-Schiotz,03S-Schiotz}, radiation damage~\cite{gao2000atomic}, dynamic fracture, and crack propagation~\cite{vashishta1996crack,vashishta1999large}, etc., the required system size typically ranges from thousands to hundreds of millions of atoms.
% Some of these problems demand time scales extending up to the microsecond and beyond, which are difficult to achieve even with MD simulations based on simple empirical force field (EFF). 
% Often processes requiring long time scales are rare events, i.e. they involve relatively fast atomic reorganizations but happen very rarely. Examples are nucleation processes and chemical reactions with large activation barriers. 
% Special simulation techniques that introduce a bias to enhance the sampling of rare events have been devised to deal with such situations~\cite{laio2002escaping,zhang2018reinforced}. 
% These techniques make rare events amenable to MD simulation but still require relatively long time scales on the order of tens or hundreds of nanoseconds.
Some of these problems demand time scales extending up to the microsecond and beyond, which is far out of the scope of AIMD. 
Although special simulation techniques that introduce a bias to enhance the sampling of the slow processes have been devised to deal with such situations~\cite{laio2002escaping,zhang2018reinforced}, they still require MD simulations of relatively long time scales on the order of tens or hundreds of nanoseconds.
Some problems demand an even higher accuracy, e.g., the so-called chemical accuracy ($\sim$1 kcal/mol), than DFT could provide, requiring more expensive methods like CCSD(T)~\cite{purvis1982CCSD}, whose computational complexity scales with the seventh power of the system size. 
%Some of the problems demand an even higher accuracy, e.g., the so-called chemical accuracy ($\sim1$ kcal/mol), for which even DFT would fail and more expensive methods like the CCSD(T)~\cite{purvis1982CCSD}, whose computational complexity scales with the seventh power of the system size, are required.
Although there have been a host of empirical force fields (EFF)-based MD schemes (see, e.g., Refs.~\cite{van2001reaxff,wang2004development,brooks2009charmm,jelinek2012modified,senftle2016reaxff}), which can easily scale up to millions,  or even trillions, of atoms,
%due to the relatively simple form of EFFs, 
their accuracy is often in question.
In particular, it has been challenging to develop EFFs for cases involving multiple elements or bond formation and cleavage, and for many practical problems there are no suitable EFFs available.
%\lz{
Recently, reactive force fields capable of modeling chemical reactions, such as the REAXFF method introduced by Goddard and collaborators ~\cite{van2001reaxff,senftle2016reaxff}, have attracted considerable attention. These methods, however, lack the generality and predictive power of DFT. 
%}\LZ{added according to Roberto's suggestion.}
Above all, there is an urgent demand in the MD community for fundamentally boosting the efficiency of AIMD while keeping its accuracy.

\subsection{Deep Potential Molecular Dynamics}

Recently, machine learning based MD (MLMD) schemes~\cite{behler2007generalized,bartok2010gaussian,chmiela2017machine,schutt2017schnet,smith2017ani,han2017deep,zhang2018deep,zhang2018end}  offer a new paradigm for boosting AIMD by means of ML-based models trained with \textit{ab initio} data.
One such model, Deep Potential (DP),
%,  trained in an end-to-end manner using \textit{ab initio} data,
has demonstrated the ability to achieve an accuracy comparable to AIMD, and an efficiency close to EFF-based MD~\cite{zhang2018deep,zhang2018end}. 
The accuracy of the DP model stems from the distinctive ability of deep neural networks (DNN) to approximate high-dimensional functions~\cite{barron1993universal,ma2019machine}, 
the proper treatment of physical requirements like symmetry constraints,
and the concurrent learning scheme that generates a compact training dataset with a guarantee of uniform accuracy within the relevant configuration space~\cite{zhang2019active}.

\begin{figure}
  \centering
  \includegraphics[width=0.45\textwidth]{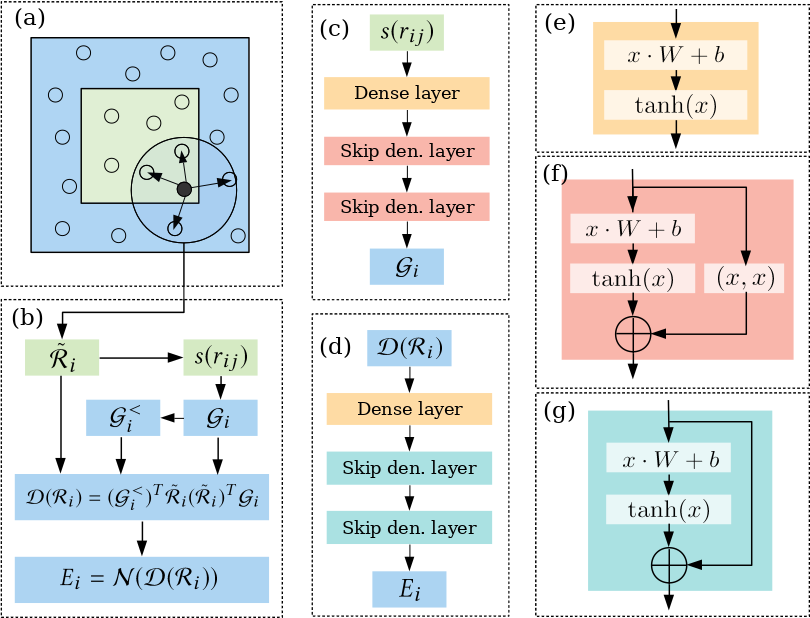}
  \caption{Schematic plot of the DP method. 
  (a) A sub-region, including the local sub-region (green) and the ghost region (blue), handled by an MPI task.
  (b) Mapping of the local environment of a single atom onto atomic energy contribution. 
  (c) Structure of the embedding net. 
  (d) Structure of the fitting net. 
  (e) Dense layer used in the embedding net. 
  (f) Skip connected dense layer used in the embedding net. 
  (g) Skip connected dense layer used in the fitting net.
  }
  \label{fig:dp}
\end{figure}

As shown in Fig.~\ref{fig:dp}, to construct a DP model,
first, 
the coordinates of atom $i$ and of its neighboring atoms are converted to the \emph{descriptors} $\mathcal D$, which encode the local atomic environment through a set of symmetry preserving features and trainable parameters. 
Next, the descriptors are passed to the \emph{fitting net}, a fully connected DNN denoted by $\mathcal N$, which outputs the atomic energy contribution $E_i$. 
Finally, the potential energy is constructed as the summation of $E_i$.
In detail, the descriptor $\mathcal D$ is the product of terms involving the \emph{environment matrix} $\tilde{\mathcal R}$, which faithfully records the relative positions of the neighbors, 
and the \emph{embedding matrix} ${\mathcal G}$, which encodes the information of the distances between atoms by a DNN named \emph{embedding net}.
% The contraction between the environment and the embedding matrices leads to an invariant representation under any permutation of atoms of the same chemical species; and the contraction between two environment matrices guarantees the invariance under rotational transformations of  the of neighboring atoms. 
The dependence of DP on the atomic coordinates is continuous to at least the 2nd order in the atomic displacements. 
% See details in Ref.~\cite{zhang2018end}.}\LZ{I add this sentence to avoid the criticism about the smoothness of DP.}
The training of the DP model has been implemented in the DeePMD-kit package~\cite{wang2018kit}. 
The typical training time spans from several hours to one week on a single GPU card, depending on the complexity of the data.

\begin{table*}[]
    \centering
    \caption{Performance of molecular dynamics simulators with \textit{ab initio} accuracy. The abbreviations Pot., TtS, LS, BP, and DP stand for potential, time-to-solution, linear scaling, Behler-Parrinello scheme, and Deep Potential, respectively. 
    In AIMD, we assume 5 electronic steps for each MD (ionic) step. 
    The time-step of water system is 0.5~fs, and that of other systems is 1~fs. 
    *The parallel efficiency does not significantly decay at the largest machine scale tested in the work, so it is highly likely that they can scale to larger machines.
    $\dagger$Vienna Scientific Cluster (VSC), an HPC system with Intel Xeon Gold 6138 CPUs.
    $\ddagger$An unknown cluster with Intel Xeon E5-2650v2 CPUs at the KISTI supercomputing center.
    **The baseline DeePMD-kit implementation.}
  \begin{tabular}{l c c c  r r r r r @{\hspace{3em}}l@{ $\times$ }l}
    \toprule
    Work & Year & Pot. & System & \#atoms & \#CPU {\small cores} & \#GPUs & Machine & \multicolumn{1}{c@{\hspace{2.5em}}}{Peak[{\small FLOPS}]} & \multicolumn{2}{@{\hspace{-2em}}r}{TtS [{\small s/step/atom}]}\\ \midrule
    Qbox~\cite{gygi2006large} & 2006 & DFT & Mo &  1K & 262K & -- & BlueGene/L & 207T & $2.8$ & $ 10^{-1}$  \\
    LS3DF~\cite{wang2008linearly} & 2008 & LS-DFT & ZnTeO &16K & 131K & --& BlueGene/P & 108T & $1.8$ &$10^{-2}$ \\
    RSDFT~\cite{hasegawa2011first} & 2011 & DFT & Si & 107K & 442K & -- & K-computer & 3.1P & $2.6$ &$ 10^0$  \\
    DFT-FE~\cite{das2019fast} & 2019 & DFT & Mg & 11K &  159K & 22.8K & Summit & 46P & $6.5$ & $ 10^{-2}$ \\
    CONQUEST~\cite{nakata2020large} & 2020 & LS-DFT & \ce{Si} & 1M & 200K & -- &  K-computer & ? & 4.0 & $10^{-3}$ \\
    Simple-NN~\cite{lee2019simple}* & 2019 & BP & \ce{SiO2} & 14K & 80 & -- & Unknown$\ddagger$ & ? & $3.6$ & $ 10^{-5}$  \\
    Singraber {\small el.al.}~\cite{singraber2019library}* & 2019 & BP & \ce{H2O} & 9K & 512 & -- & VSC$\dagger$ & ? & $1.3$ & $ 10^{-6}$  \\
    Baseline~\cite{wang2018kit}** & 2018 & DP & \ce{H2O} & 25K & 1 & 1 & Summit & -- & $5.6$ & $10^{-5}$ \\
      \midrule
%    This work & DP & \ce{H2O} & -- & 12,582K-402,653K & 4.6K & 27.3K & Summit & \WH{fill here} & \WH{fill here} \\
    This work (double)     & 2020 & DP & \ce{H2O} & 679M & 27.3K & 27.3K  & Summit & 80P  & $3.0$ & $ 10^{-10}$   \\
    This work (mixed-half) & 2020 & DP & \ce{H2O} & 679M & 27.3K & 27.3K  & Summit & 212P & $1.1$ & $ 10^{-10}$   \\
    This work (double)     & 2020 & DP & Cu & 127M & 27.3K & 27.3K & Summit & 91P  & $8.1$ & $ 10^{-10}$  \\   
    This work (mixed-half) & 2020 & DP & Cu & 127M & 27.3K & 27.3K & Summit & 275P & $2.7$ & $ 10^{-10}$  \\     
%    This work & DP & Cu & -- & 7,077K-127,246K & 27.3K & 27.3K & Summit & \WH{fill here}  & \WH{fill here}  \\
    \bottomrule
    \end{tabular}
    \label{tab:soa}
\end{table*}

Deep Potential Molecular Dynamics (DeePMD) has greatly boosted the time and size scales accessible by AIMD without loss of \textit{ab initio} accuracy.
To date, DeePMD has been used to model
%study problems like first-order phase transitions~\cite{bonati2018silicon}, infrared spectroscopy and Raman spectroscopy~\cite{zhang2020dw,sommers2020raman}, nuclear quantum effects~\cite{ko2019isotope}, and 
various phenomena in physical chemistry~\cite{andrade2020free,zeng2019neural,chen2018deep,zhang2020dw,ko2019isotope} and materials sciences~\cite{dai2020theoretical,marcolongo2019simulating,wang2019deep,20JPCM-Liu,bourgeois2020transforming,niu2020ab}.
For example, in a recent work~\cite{andrade2020free}, DeePMD was used to simulate the TiO${}_2$-water interface,
providing a microscopic answer to an unsolved question in surface chemistry: do water molecules dissociate or remain intact at the interface between the liquid and TiO${}_2$?
%\RC: water adsorption in dissociated or undissociated form is not an unsolved question for the surface in vacuum, the issue is what happens at the interface between the liquid and TiO2
In another recent work~\cite{bourgeois2020transforming}, DeePMD  was used in combination with experiments to show the mechanism behind the nucleation of strengthening precipitates in high-strength lightweight aluminium alloys.
These examples are challenging for AIMD due to the spatial and temporal limits of this approach.
They are also difficult, if not impossible, for EFF-based MD schemes, due to the limited capability of the relatively simple form of the potential energy function they adopt.

\section{Current state of the art}
%\WL{GB requirement for this section: quantitative discussion of current SoA, with emphasis on performance-related aspects  (1 p max)}

%\LZ{I realized from Lin's comment and 19 GB paper that although we could list a lot of advances in AIMD and MLMD, we shouldn't call all of them the state-of-the-art. Instead, we may set as the target really large-scale, long-time, and ab-initio accurate simulation.  Then we review conventional AIMD methods and say that there will be a long way to go if we stick to these methods; Then we review MLMD methods and say that no one is suitable for modern heterogeneous supercomputers and few efforts have been spent from the HPC aspect; Finally we introduce the baseline implementation of DeePMD-kit.}

An important goal of molecular simulation is to model with \textit{ab initio} accuracy realistic processes that involve hundreds of millions of atoms.
%Although some empirical force field (EFF)-based MD schemes have scaled up to trillions of atoms~\cite{tchipev2019twetris} or reached simulation time-scales of milliseconds~\cite{shaw2014anton}, their accuracy is often in question and feasible systems have been limited
%\LL{ this sentence needs some modification}. \WH{We may not need to mention EFFMD in this review section, because it has been excluded by the last paragraph of 3.1. Here we need to define what is the "target system" (hundreds of millions of atoms in some applications like ...) and by the table present the discrepency between the target and the existing AIMD methods.}
%To achieve this goal, there have been tremendous efforts made for boosting AIMD without loss of its accuracy.
To achieve this goal, major efforts have been made to boost AIMD without loss of accuracy.
Some examples are QBox~\cite{gygi2006large}, LD3DF~\cite{wang2008linearly}, RSDFT~\cite{hasegawa2011first}, DFT-FE~\cite{das2019fast}, and CONQUEST~\cite{nakata2020large}. 
Their performances are summarized in Table~\ref{tab:soa}, where the system size, the peak performance, the time-to-solution, etc., are provided. 
We observe that it is challenging for conventional DFT-based AIMD schemes to overcome the 
%limit of the cost even by the fastest HPCs nowadays. 
cost limits even with the fastest available HPCs.
As a rough estimate, 
assuming that future HPC performance will continue to improve at the same pace as in the past fourteen years, it would take several decades to be able to model the target size and time scales of interest with conventional AIMD techniques.

The MLMD schemes mentioned in the last section offer a chance to bypass the conventional AIMD methods without losing their accuracy.
Representative examples are the Behler-Parrinello scheme~\cite{behler2007generalized}, the Gaussian approximation potential~\cite{bartok2010gaussian,bartok2013representing},
SchNet~\cite{schutt2017schnet}, and the Deep Potential method~\cite{han2017deep,zhang2018deep}. 
Up to now, most attentions of the community have been devoted to improving the representability and transferability of the machine learning schemes, and to solving scientific problems that do not really require very large-scale MD simulations. 
Efforts on implementation and optimization with an HPC perspective have remained at an early stage. 
Some open-source packages for the MLMD schemes have been released: the QUantum mechanics and Interatomic Potentials (QUIP)~\cite{quip}, Amp~\cite{khorshidi2016amp}, DeePMD-kit~\cite{wang2018kit}, TensorMol~\cite{yao2018tensormol}, SIMPLE-NN~\cite{lee2019simple}, PES-Learn~\cite{abbott2019pes}, and a library-based LAMMPS implementation of neural network potential~\cite{singraber2019library}. 
The performance reported in these works, if any, is summarized in Table ~\ref{tab:soa}.
It is observed that existing implementations of MLMD are basically for desktop GPU workstations or CPU-only clusters.
None of them can fully utilize the computational power offered by the accelerators on modern heterogeneous supercomputers.

Of particular relevance to our work is the DeePMD scheme, which has been implemented in an open-source package called DeePMD-kit~\cite{wang2018kit}.
DeePMD-kit is built on the MD platform LAMMPS \cite{plimpton1995fast} and the deep learning platform TensorFlow~\cite{abadi2016tensorflow}. 
By interfacing the DP model with LAMMPS, which maintains the atomic information and integrates the equations of motion, the key function of DeePMD-kit is to implement the calculation of atomic energies and forces predicted by the DP model.
With TensorFlow, a versatile tool box for deep learning, the embedding matrix, the descriptors, and the atomic energy are implemented by standard operators built in TensorFlow. 
Moreover, TensorFlow provides GPU support for its standard operators, thus the corresponding calculations in DeePMD-kit are easily accelerated with GPU by linking to the GPU TensorFlow library. 
Unfortunately, the implementation of DeePMD-kit cannot fully utilize the computational power of modern heterogeneous supercomputers like Summit, due to the following restrictions: 
(1) The code is designed on single node with only single GPU serial or multi-CPU OpenMP parallelism~\cite{wang2018kit}. 
%The task of computing the forces in the system was not spatially partitioned into sub-tasks that are assigned to sub-regions maintained by LAMMPS processors. 
(2) The customized TensorFlow operators introduced for the environment matrix, force, and virial are implemented only on CPUs.  
% The two sorting operations needed by the environment matrix are efficient on CPUs, but is highly non-trivial on GPUs. 
(3) The size of the DNN used by DP models is typically smaller than the sizes adopted in normal deep learning applications like pattern detection and language processing, which implies that each individual step of a computationally intensive operation is also smaller in DP applications. 
In this context, the memory bandwidth and latency become obstacles to improving the computational efficiency of the DeePMD-kit package.
To summarize, large-scale DeePMD simulations with \textit{ab initio} accuracy have been only conceptually proved to be possible, but have never been made practically accessible by a code optimized for modern heterogeneous HPCs, from both algorithmic and implementation perspectives. 

Above all, to the best knowledge of the authors, efficient MD simulation of 100 million atoms with \textit{ab initio} accuracy has never been demonstrated with AIMD or MLMD schemes.
We believe that to make this goal a routine procedure, we need to pursue integration of physics-based modeling, machine learning, and efficient implementation on the next-generation computational platforms.
In the following sections,  we shall adopt the serial DeePMD-kit~\cite{wang2018kit} as the baseline DeePMD implementation
and demonstrate how its performance can be greatly boosted on Summit.

\section{Innovations}\label{sec:innovation}
%\WL{GB requirement for this section:what the innovations are and how they were achieved (2 pp max)}

% \WH{The DP model itself should not be the innovation of this work. The innovation is to implement the DP model for heterogeneous supercomputer. Therefore, I revised this section.} \WL{yes, I agree.}
%\LL{ This is a key section and I suggest that we could consider rewrite it as follows. Basically it should be of higher level than the current version.}

%\LL{ I wonder whether there is a way to illustrate more clearly that these modifications are novel from a computer science perspective.  Is this implementing standard practices in a new context, or some key steps (such as reformatting the neighbor list / mixed precision treatment of selected components) are novel and can be used in other contexts?} 

%\WL{By the order of novelty, I will say we did some nice work in the NN, by replacing MATMUL and SUM with GEMM(althoug we need to avoid saying that google did some stupid job.). The compression of the neighbor list element is one trick that applies to all other NNMD code, it is nice. Mixed precision is good, although this is quite common. So are the other techiniques, such as SoA to AoS. All in all, we did not do a perfectly fancy job in CS, like what Torsten did last year, but what we do works for the NNMD.  }

\subsection{Summary of contributions}

\begin{figure*}[t]
  \centering
  \begin{subfigure}{0.32\textwidth}
    \includegraphics[width=\linewidth]{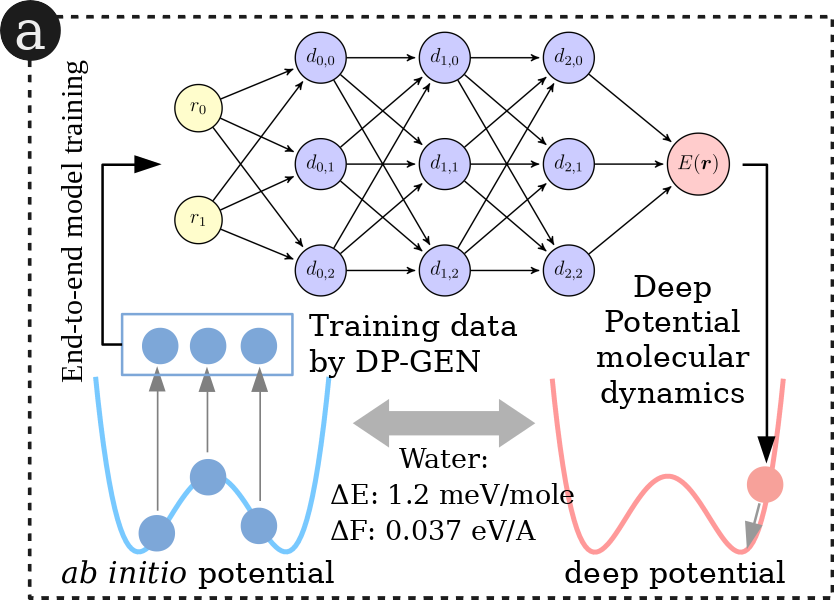} \\
    \includegraphics[width=\linewidth]{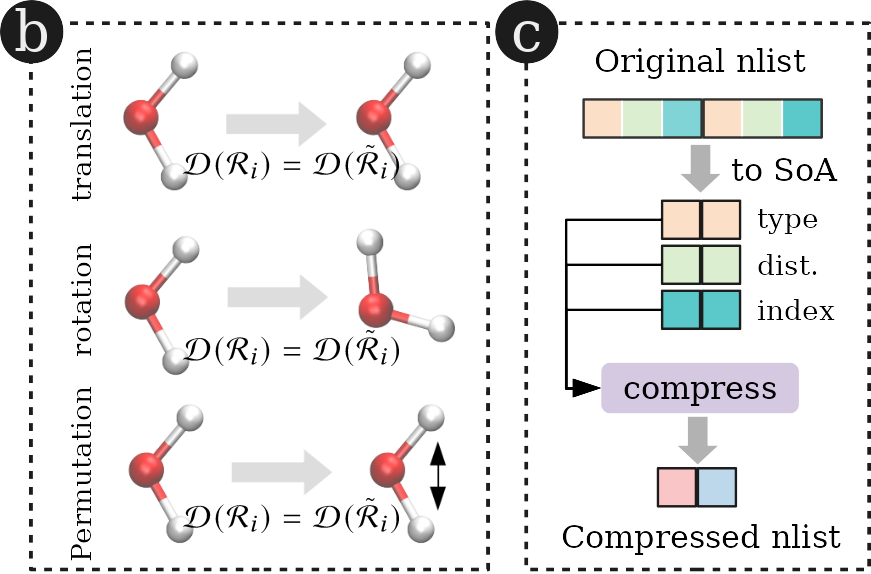} \\
  \end{subfigure}\hfil
  \begin{subfigure}{0.32\textwidth}
    \includegraphics[width=\linewidth]{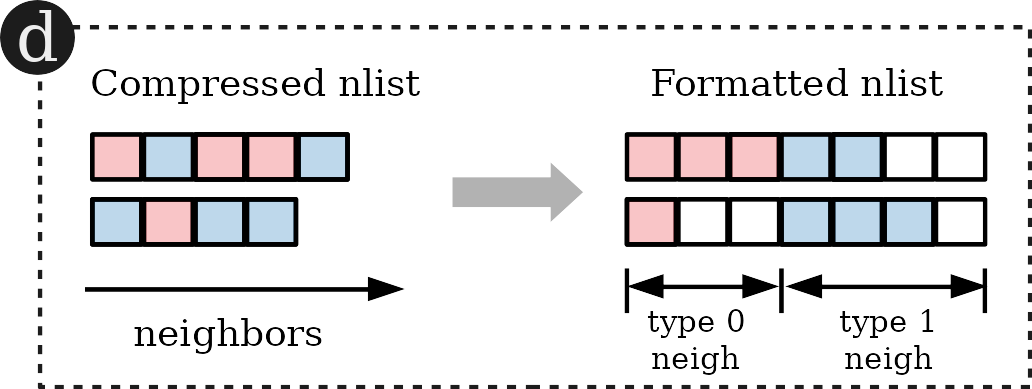} \\
    \includegraphics[width=\linewidth]{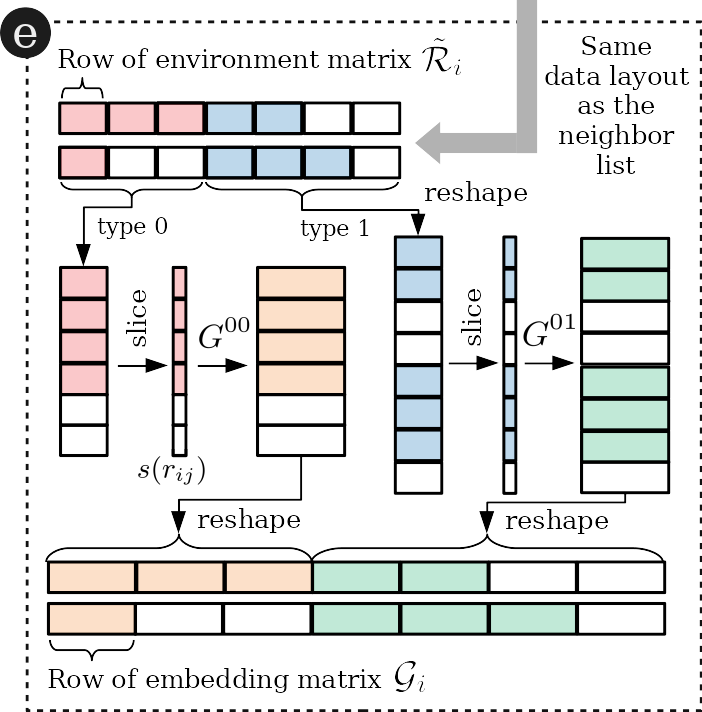} \\
  \end{subfigure}\hfil
  \begin{subfigure}{0.32\textwidth}
    \includegraphics[width=\linewidth]{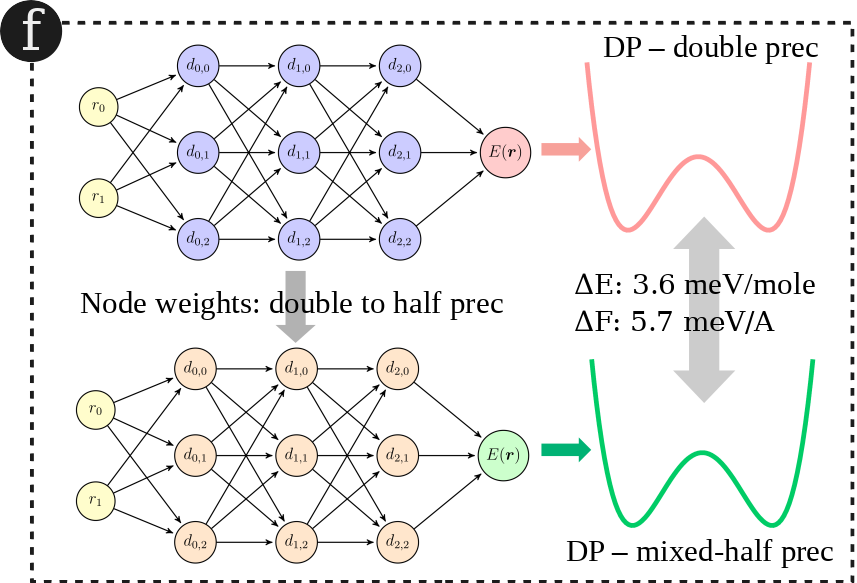} \\
    \includegraphics[width=\linewidth]{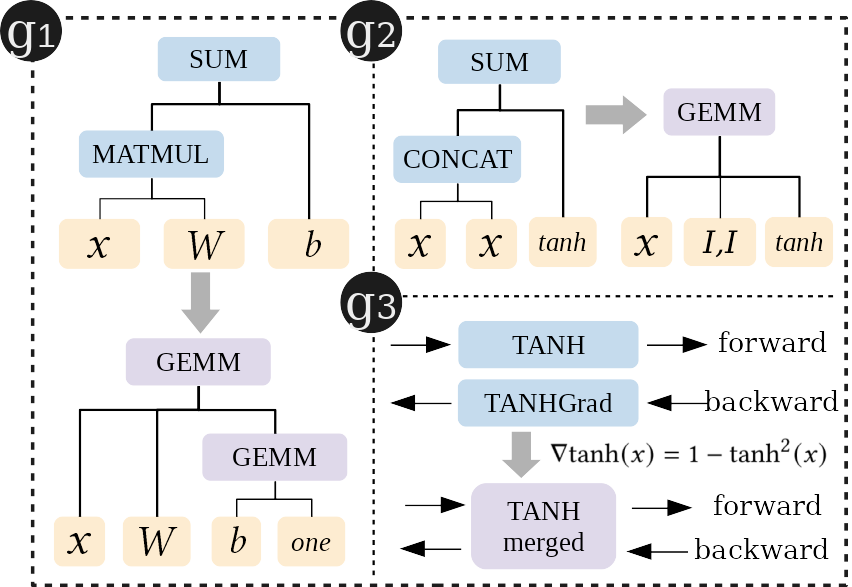} \\
  \end{subfigure}
  \vskip -.1cm
  \caption{Key steps in the optimized DeePMD-kit, taking water as an example.}\label{fig:opt}
\end{figure*}

Our major contribution is a highly efficient and highly scalable method for performing MD simulation with \textit{ab initio} accuracy. 
This is achieved by combining the unprecedented representation capability of the DP model (Figs.~\ref{fig:opt}~(a)-(b)), and a highly scalable and fine-tuned implementation on heterogeneous GPU architectures (Figs.~\ref{fig:opt}~(c)-(g)). 
%One key factor is that in terms of computational intensity, DPMD is between standard potential energy models and AIMD simulation. \WH{The purpose of this sentence is not clear for me. The term "standard potential" was not introduced.} Therefore it is crucial to properly organize the computational tasks and to reduce the cost of data communication. \WH{@LL, it seems that this sentence is not the consequence of the previous one. We may not want to use "therefore".  }
%Since deep learning based molecular dynamics simulation is still at its infancy, we expect that our optimization strategy can be useful for future deep learning models for molecular dynamics simulation. 
The resulting optimized DeePMD-kit scales almost perfectly up to 4,560 computing nodes on Summit for a copper system of 127,401,984 atoms,
reaching $91$ PFLOPS in double precision, and $162$ and 275 PFLOPS in mixed single and mixed half precision, respectively. 
The corresponding time-to-solution is {$34$} milliseconds per MD step {with mixed half precision}, outperforming existing work by more than three orders of magnitude and enabling nanosecond simulation within {$10$} hours. 
%Our time-to-solution outperforms existing work by more than three orders of magnitude, and brings the molecular dynamics simulation with \textit{ab initio} accuracy into a new era.
%\LZ{Roberto: use a less bombastic language }

\subsection{Algorithmic innovation}

To effectively harness the computing power offered by the heterogeneous system architecture of Summit, our goal is to migrate to GPUs almost all computational tasks and a significant amount of communication tasks.
%(via CUDA-aware MPI). 
Due to the relatively limited size of the computational granularity in the DP model, a straightforward GPU implementation encounters many bottlenecks and is thus not efficient. 
As such, our main algorithmic innovations are the following:

\begin{itemize}
    %\item We show algorithmic innovation in DeePMD-kit to approximate \textit{ab initio} MD with high accuracy. \LL{A bit more specific. Discuss the choice of data layout that allows us to significantly increase the computational granularity.} 
    \item  We increase the computational granularity of DeePMD  by introducing a new data layout for the neighbor list that avoids branching in the computation of the  embedding matrix.
    %\item We redesign the data structure of neighbor list, and optimize the customized TensorFlow operators on GPU to achieve an overall speedup factor of $6.2$.  \WH{Two innovations regarding the neighbor list: 1, Algorithmic: the new structure of the neighbor list (avoiding branching in the computation of embedding matrix, ciritcal to GPU), 2, Implementation: efficient implementation of this data structure on GPU (optimizing the customized TF ops).} \WL{got u, but not sure how to write, keep the old one for now}
    \item 
    % Based on the new data structure, we optimize computationally intensive tasks in the customized TensorFlow operators.
    %Based on the new data structure, we develop the GPU algorithms for the customized TensorFlow operators and optimize computationally intensive tasks. 
    The elements in the new data structure of the neighbor list are compressed into 64-bit integers for more efficient GPU optimization of the customized TensorFlow operators.
%    \WH{@Weile, please check this sentence.}\WL{can we say something like this: `` The new data structure of the neighbor list is compressed into a 64-bit integer for more efficient optimization of the customized TensorFlow operators.''}
    \item We develop mixed-precision computation for the DP model. Computationally intensive tasks are performed with single or half precision without reducing the accuracy of the physical observables.
    %\item We parallelize the DeePMD-kit code on the heterogeneous supercomputer Summit, and optimize the MPI communication to enhance its scalability. 
    %\item We develop the task partitioning algorithm for DP, successfully parallelize the baseline DeePMD-kit package, and optimize its MPI communication on Summit. 
\end{itemize}

\subsubsection{Increasing  computational granularity}\label{sec:granularity}

The novelty of the DP model lies in its ability to automatically generate a set of symmetry-preserving descriptors $\mathcal D$ through the embedding net (Figs.~\ref{fig:dp}~(b) and (c)) from the local environment of each atom described by the environment matrix $\mathcal R_i$. 
By using roughly the same set of hyper-parameters, DP can fit the data for almost all tested systems. 
Compared to other methods with fixed feature sets, DP is more versatile when facing complex data, e.g., multi-component systems, chemical reactions, etc. 
Since %translational, rotational and permutational invariance of the physical system is 
important symmetries are strictly preserved in $\mathcal D$ (see Fig.~\ref{fig:opt}~(b)), a fitting network of three layers (see Fig.~\ref{fig:dp}~(d)) is enough to produce results with high fidelity. 

The most computationally intensive part of the DP model is the embedding matrix. 
The pattern of the computation 
%is defined by the neighbor list, i.e.~the order in which the rows of the embedding matrix is computed and stored 
is defined by the order of neighbors recorded in the neighbor list. 
% \LZ{@weile, I commented out some redundant words, please check.}
We notice that since the descriptors are permutationally invariant (Fig.~\ref{fig:opt}~(b)), neighbor lists with different orders are equivalent in terms of accuracy. 
By taking advantage of this observation, we redesign the data layout of the neighbor list by sorting the neighbors according to their type,  and, within each type, we sort the neighbors by their relative distance. 
The neighbors of the same type are padded to the cut-off number of neighbors corresponding to that type (Fig.~\ref{fig:opt}~(d)).
The first sorting (according to neighbor types) and the padding align the neighbors with the same type, so the conditional branching according to the type of the neighbors in the embedding matrix computation is avoided (see Fig.~\ref{fig:opt}~(e)). 
This greatly increases the computational granularity, a critical component for taking advantage of the computational power offered by GPUs. 
The second sorting always selects the neighbors in the list according to their distance from the central atom. In this way, if the number of neighbors
occasionally fluctuates beyond $N_{max}$, the cut-off number of neighbors defined in the padding step, only information on the nearest neighbors up to 
$N_{max}$ is retained, avoiding the unphysical phenomena that would occur if close neighbors were neglected.   
%the neighbor list, so that unphysical phenomena are avoided when the number of neighbors of a certain type occasionally fluctuates beyond the cut-off %number of neighbors defined in the padding step.
% \LZ{To Roberto: this sentence means that, in some rarely happened cases, during a simulation, the number of neighbors of a certain atom may exceed the predefined maximum value $N_{max}$. In this case, we keep the information of the nearest $N_{max}$ neighbors.}
% \LZ{@Han, I commented out the last sentence here since I think we don't need to enter into such details given that we have defined the ``maximally possible'' number.}

%\WL{ I copied the following part from SC'20 paper, need revision to serve this paper}
%The central quantity of an MD simulation is the potential energy surface $E$, a function of the atomic coordinates $(r_1, \dots r_N) \in \mathbb R^{3N}$. The DP model expresses $E$ as a sum of atomic contributions, i.e., ~$E = \sum_i E_i$. The contribution $E_i$ from the atom $i$ depends only on $\mathcal R_i$, the local environment of $i$: $\mathcal R_i=\{r_{ij} : j \in L(i)\}$, where $r_{ij} = r_j - r_i$. Here the neighbor index set $L(i)$ is defined by $\{j: \vert r_{ij} \vert \leq r_c\}$, and $r_c$ is a predefined cutoff radius. In the DP model, $\mathcal R_i$ is first mapped via an embedding net onto a symmetry-preserving descriptor $\mathcal D$, and then $\mathcal D$ is mapped via a fitting network $\mathcal N$ to give $E_i$, i.e.,
%\begin{align}\label{eq:ei}
%  E_i = \mathcal N (\mathcal D (\mathcal R_i)).
%\end{align} 

\subsubsection{Optimization of Customized TensorFlow Operators}\label{sec:gpu_optimization}

In this part, we present the optimization of the customized TensorFlow operators, which take more than $84\%$ of the total computational cost in the baseline DeePMD-kit. 
We start from formatting the neighbor list, whose data layout is crucial and discussed in Sec.~\ref{sec:granularity}. 
Each element of the neighbor list is a structure with 3 data items: the atomic type $\alpha(j)$, the atomic distance $\vert r_{ij}\vert$, and the atomic index $j$ (Fig.~\ref{fig:opt}~(c)). 
In the formatting process, the neighbor list is sorted first based on the atomic type, then based on the atomic distance $\vert r_{ij}\vert$. 

The AoS (Array of structures) data layout of the neighbor list makes it impossible for efficient memory access on GPU because of memory coalescing problems.
One common practice in GPU optimization is to switch from AoS to SoA (Structure of arrays).
However, in DeePMD-kit, we propose an even more efficient way of storing the neighbor list by compressing each element of the neighbor list into a 64-bit unsigned integer (Fig.~\ref{fig:opt}~(c)) with the following equation:
$
\alpha(j) \times 10^{16} + \lfloor \vert r_{ij}\vert \times 10^{8} \rfloor \times 10^{6} + j
$.
The 20 decimal digits of the 64-bit unsigned integer are divided into 3 parts to store the one element of neighbor list: 
4 digits for the atomic type, 10 digits for the atomic distance, and 6 digits for the atomic index.
% \LuDh{Atomic index store has been changed form 5 digits to 6 digits. For some cases, for example, number of atoms reaches 60,000 per GPU, the index of the neighbor could be greater than 100,000. so the last part use 6 digits to store the index(include ghost neighbors) safely. Our code is now using this new formula compared with the April version, fix a bug here}
% 4 decimals are used to store the atomic type  ($\alpha(j)$), 10 decimals are used to store the atomic distance ($\vert r_{ij}\vert$), 5 decimals are used to store the atomic index ($j$). 
The range of all the three parts are carefully chosen and are rarely exceeded in typical DeePMD simulations. 
Both the compression before sorting and the decompression after sorting are accelerated via CUDA customized kernels, so that the corresponding computational time is negligible.
Sorting the compressed neighbor list reduces the number of comparisons by half with no impact on the accuracy of the algorithm, and is carried out by calling the NVIDIA CUB library, which provides the state-of-the-art and reusable software components for each layer of the CUDA programming model, including block-wide sorting. 

According to Amdahl's law, an ideal overall speedup can only be achieved by accelerating all calculations. 
In our implementation, all customized TensorFlow operators, Environment, ProdForce, and ProdViral, which compute the environment matrix, force, and the virial, respectively, are migrated and optimized on the GPU. 
%Note that in the optimization, 
In particular, a fine-grained parallelism is utilized to exploit the computing power of the GPU. 

Now that all computationally intensive tasks are carried out by the GPU, we further reduce the time for GPU memory allocation by allocating a trunk of GPU memory at the initialization stage, and re-using the GPU memory throughout the MD simulation. 
The CPU-GPU memory copy operations are also optimized to eliminate non-essential data transfer processes.

\subsubsection{Mixed-precision computation} %\WL{no mixed precision MPI is used. and shall we consider moving this part into section 5.3? reason is: we only use mixed precision in the Neural network.} \WH{agree with WL}

The approximation property of the DNN-based DP model provides us with an opportunity for mixed-precision calculations. 
In the optimized code, different levels of mixed precision are tested, and we find that 
two prescriptions of mixed precision are of satisfactory stability and accuracy. 
Both of them use double precision for atomic positions and the environment matrix construction.
%The scheme named {MIX-32} casts the environment matrix to single precision (32-bit) and uses single precision for the following floating point operations. 
In the MIX-32 scheme, all parameters of the embedding net and fitting net are stored in single precision (32-bit). The environment matrix is converted from double precision to single precision, then all the arithmetic operations of the embedding net and the fitting net are performed in single precision. 
{In the MIX-16 scheme, the parameters of the embedding net and the first two fitting net layers are stored in half precision (16-bit). The environment matrix is cast to half precision and then fed to the embedding net. In each embedding net layer and the first two fitting net layers, the GEMM operations are performed using Tensor Cores on V100 GPU with accumulations in single precision, except for those in the first embedding net layer that do not meet the size requirements for using Tensor Cores. All other floating point operations, such as TANH and TANHGrad, are conducted in single precision due to accuracy considerations. 
The data are cast to half precision before writing the global memory. }
% In the MIX-16 scheme, the parameters of the embedding net and fitting net are stored in half precision (16-bit), however, most arithmetic operations are still performed using single precision for accuracy purposes. For examples, the GEMM operations are performed using Tensor Cores on V100 GPU with accumulations in single precision, and all other floating point operations such as TANH and TANHGrad are also conducted in single precision. 
%The scheme named MIX-16 casts the environment matrix to half precision (16-bit), and stores all input and output of the successive hidden-layers in half precision.
%All GEMM operations in MIX-16 are performed on Tensor Cores on V100 GPU with accumulations in single precision, and other floating point operations are conducted in single precision.
% \LuDh{"All GEMMs in MIX-16 are performed in half precision" is not accurate here. The cutlass and default cublas take input data in half floating-point precision, perform matrix multiplication in half precision and the accumulation in single precision. like, Matrix multiply-add operation: F32 = F16 * F16 + F32}
% It is noted that the network paramters $W$ and $b$ are stored in single precision and are casted to half precision before GEMMs.
%\wh{The last layer of the fitting net has the same construction as the fitting net layers in the MIX-32 scheme, i.e.~the data storage and floating point operations are in single precision.We remark that keeping the last layer of fitting net in single precision is critical to the overall accuracy of the MIX-16 scheme. }
Note that in the last layer of the fitting net, both data storage and arithmetic operations are kept in single precision, which is critical to the accuracy of the MIX-16 scheme. 
% The only exception in the MIX-16 scheme is the last layer of the fitting net, wherein the data storage and floating point operations are in single precision. 
% %single precision implementation in the TensorFlow is stable and accurate. 
% a stable and accurate prescription is to 
% %Notice that here by single precision implementation we mean to 
% use single precision for all the network parameters, but use double precision for 
% other parts like the atomic positions and energies.
% %are still in double precision.
%\LZ{We may not need to define ``single precision implementation" and then explain it in the above two sentences.}
% \wh{In detail, the environment matrix, constructed in double precision, is converted to single precision (MIX-32) or half precision (MIX-16) and then is used to predict atomic energy and force by DNNs in single precision (MIX-32) or half precision (MIX-16, last layer in single precision)}. \wl{ Note that in the MIX-16 code, all floating point operations are evaluated in FP32 precision using Tensor Cores, and then the weight $W$ and bias $b$ are stored in FP16 precision. }
% \LL{ what does ``all numerical calculations'' mean?} \WL{changed to ``floating point operations'' }
Finally, the outputs of the fitting net of MIX-32 and MIX-16 are converted back to double precision, and the total energy of the system is reduced from the atomic contributions.
The model parameters $W$ and $b$ are trained in double precision, and cast to single and half precision in the MIX-32 and MIX-16 schemes, respectively.

We compare the mixed-precision schemes
% \WH{in the following we use "model" that is inconsistent with "scheme"}
with the double precision by using a typical configuration of a water system composed of 512 molecules. 
With MIX-32 we observe a deviation of $5.2\times10^{-6}$~eV (normalized by the number of molecules) in the energy prediction and a root mean square deviation of $2.5\times 10^{-6}$~eV/\AA\ in the force prediction, which indicates an excellent agreement with the double precision scheme.
With MIX-16 we observe a deviation of $3.6\times10^{-3}$~eV (normalized by number of molecules) in the energy prediction and a root mean square deviation of $5.7\times 10^{-3}$~eV/\AA\ in the force prediction.
The deviation in the force prediction is significantly smaller than the training error ($\sim$4$\times$$10^{-2}$~eV/\AA). The deviation in energy prediction is comparable to the training error, but is already much smaller than the chemical accuracy ($\sim$4$\times 10^{-2}$ eV/molecule). The accuracy of the mixed-precision schemes in predicting physical observables is further validated in Sec.~\ref{sec:perf-mixed}.
% \LZ{The last two sentences sound redundant. How about the following: ``Further validations on the accuracy of the mixed-precision schemes are presented in Sec.~\ref{sec:perf-mixed}.'' 
% I'm not sure whether this is the best place to present these numbers. Shouldn't we merge these results in Table 4 and introduce them in Sec.~\ref{sec:perf-mixed}?} \WH{Not sure, we can discuss on phone}
% % Since these deviations are less than the training error of the double precision model, the mixed precision implementation is of satisfactory accuracy. 
% In terms of speed and GPU memory requirement, the \wh{MIX-32 scheme} is $\sim$1.7 times faster than and consumes $\sim$50\% less memory than the double precision version, \wh{while the MIX-32 scheme is $\sim$xxx times faster than and consumes $\sim$75\% less memory than the double precision version.}
% % We remark that although half precision is more power efficient on the NVIDIA V100 GPU than single precision ($120$ TFLOPS against $14$ TFLOPS), our tests show that,  due to the limited representation range with $16$ binary bits, the corresponding DP model cannot preserve the required accuracy of the energy and forces.
% %for physical simulations.
% % Therefore, at this moment, we cannot take advantage of the merits of the half-precision implementation.

\subsection{Neural Network Innovation}\label{sec:nn} 

After optimizing customized TensorFlow operators (Sec.~\ref{sec:gpu_optimization}), the remaining computational cost is dominated by standard TensorFlow operators. 
The floating point operations are dominated by operators like {MATMUL} (matrix-matrix multiplication) and {TANH} (activation function).  
Other operators such as {CONCAT} (matrices concatenation) and {SUM} (matrix addition) are bandwidth intensive and 
{cost few floating point operations.}
We find that many operations in DeePMD-kit involve matrix-matrix multiplication of tall and skinny matrices. 
This leads to particularly large overheads in the operations 
%such as {SUM} operation, 
like {SUM}, %and hence 
so standard TensorFlow operators are not optimized to treat such matrices efficiently. 
Through detailed performance profiling, we redesign several operations in the execution graph of TensorFlow.
Although these are tailored operations designed to improve the efficiency of DeePMD-kit,  similar strategies should be useful in other machine learning applications, particularly those integrated with physical modeling.

%In the execution graph of the TensorFlow, we notice that standard operators like {MATMUL}, {SUM}, {TANH} and {CONCAT} are used to implement operations like matrix-matrix multiplication, summation of matrices, activation function, and concatenation.  The floating point operations are mainly contributed by operators like {MATMUL} and {TANH}. 
%Other operators such as CONCAT and SUM are bandwidth intensive and have little floating point operations. 
%To improve the efficiency of the standard TensorFlow operators, we introduced the following optimization techniques:  

\subsubsection{ Replace MATMUL and SUM Operators with GEMM}\label{subsec:gemm}

In the standard TensorFlow execution graph, the operation $x \cdot W +b$ (see Fig.~\ref{fig:dp} (e-g)) is implemented with two separate operators: MATMUL and SUM.  
For example, for the oxygen-hydrogen pairs in a water system with 4,096 molecules, {MATMUL} in the last layer of the embedding net multiplies $x$ of size %\wl{376,832} by \wl{64} \LZ{$376,832\times64$?} with $W$ of size \wl{64} by \wl{128}~\LZ{$64\times128$?}. %
786,432 $\times$ {64} with $W$ of size ${64} \times {128}$. 
Then the SUM operator adds the bias $b$ to each row of $x\cdot W$.
In many data-driven applications the sizes of matrices $x$ and $W$ are large enough so that the overhead of the {SUM} is negligible compared to that of the MATMUL operator. 
However, in the case of DeePMD, the second dimension of $x$ and the size of $W$ are relatively small, so the cost of SUM becomes important.
In the optimized computational graph, we replace the MATMUL and SUM operators with a single CUBLAS GEMM call.
%($C = \alpha A \times B + \beta C$).
%, since bias $b$ is not a matrix but a vector. 
Note that the vector $b$ is converted to a matrix before SUM by right multiplying with the transpose of vector \textit{one} (Fig.~\ref{fig:opt}~(g1)). 

%\WL{We remark that the replacement works well for DeePMD-kit is because DeePMD-kit has tall skinny matrix-matrix multiplication, which leads to bigger SUM overhead and is not common in other deep learning applications.} 
%\WH{@weile Repeating with what I wrote? please keep only one of them} \WL{yes, just explicitly say TensorFlow is good.}
%\WH{@Weile, check if the rephrasing works} \WL{works perfect!}

% We notice that the  matrix-matrix multiplication and summation operations are treated as two separated operators for evaluating $x \cdot W +b$ in the TensorFlow execution graph. 
% For example, in the case of a system with $4096$ water molecules, the MATMUL operator is invoked to calculate $x \cdot W$, where $x$ is the oxygen-hydrogen environment matrix of size $376,832$ by $50$ and $w$ is the weight matrix of size $50$ by $100$.  
% Then the SUM operator is called to add the bias $b$ to the resulting matrix of $x \cdot W$. 
% To improve the efficiency of these two operators, we replace the MATMUL and SUM operators with a single CUBLAS GEMM call ($C = \alpha A \times B + \beta C$). 
% We remark that in the optimized DeePMD-kit, $b$ is converted from a vector to a matrix by simply multiplying with a transpose of vector $one$. 

\subsubsection{ Replace CONCAT and SUM Operators with GEMM} 

In the standard TensorFlow computational graph, the operation $(x,x) + ...$ (see Fig.~\ref{fig:dp} (f)) is implemented by a CONCAT operator that concatenates two $x$s to form $(x,x)$ and a SUM operator that adds $(x,x)$ to the output of the TANH operator. 
We optimize this operation by replacing CONCAT with a matrix-matrix multiplication  $(x, x) \rightarrow x\times(I,I)$, and merging this multiplication with SUM to form a CUBLAS GEMM call (Fig.~\ref{fig:opt}~(g2)).
We observe that the multiplication is only marginally faster than CONCAT, and the benefit comes from the merging of the SUM. 

\subsubsection{CUDA kernel fusion for the TANH and TANHGrad} 
% We also optimize the TANHGrad and TANH operators by kernel fusion. 
TANH is the activation function (see Fig.~\ref{fig:dp} (e-g)), while TANHGrad (not explicitly shown in Fig.~\ref{fig:dp}) is the derivative of the output of TANH w.r.t the input for backward propagation. 
% of the computational graph.  
We need both TANH and TANHGrad in each MD step to evaluate the  forces.
We observe that the derivative of $\tanh(x)$ is also a function of $\tanh(x)$, i.e.~$\nabla {\tanh(x)} = 1 - {\tanh^2(x)}$. 
Thus, in the optimized DeePMD-kit, both TANH and TANHGrad operators are implemented in one CUDA customized kernel to save computational time (Fig.~\ref{fig:opt}~(g3)). 
Since the GPU memory of the TANHGrad is allocated in the forward propagation, this optimization is essentially trading space for time. 

\subsection{Reducing MPI  communication bottlenecks}

%\WH{@Weile, please check if this innovation in algorithm is too trivial for computer scientists}\WL{just one suggestion: I think it would be easier for the readers from CS background to understand the paper if we address that the TensorFlow operators(both customized and standard) are the most computational intensive parts. I think they account for more than $92\%$ of the total time. } 
%\WH{@Weili, Please see the revision in Sec.5 before 5.1.}

Despite the multi-body nature of DP, 
due to its force decomposition scheme, we can adopt for DP
%the DP, which is multi-body in nature can be adopted to 
the same parallelization scheme of the EFFs implemented in LAMMPS (Fig. 1 (a)).
% The only difference is that the Newton’s action-reaction law that is usually used to save computation and communication costs for classical force fields does not apply to DP. 
The computation of EFFs in LAMMPS is replaced by the computation of DP, and LAMMPS is also used to maintain the spacial partitioning of the system and all the communications between sub-regions.

There are mainly two types of MPI communications in each DeePMD step: the communication of the ghost region between adjacent MPI tasks and the global reduction for the physical properties. 
In our implementation, we optimize the communication of the ghost region using the CUDA-aware IBM Spectrum MPI, since it resides on the GPU in the calculation.
%Then the corresponding neighbor list is constructed directly on the GPU, and simultaneously passed to the customized TensorFlow operators without CPU-GPU memory copy operations for generating the environment matrix. 
When the output information is required, \textit{MPI\_Allreduce} operations across all MPI tasks are performed to collect physical properties, such as total energy, pressure, etc..
Although each of these physical properties is only one double precision number and the corresponding \textit{MPI\_Allreduce} operation is latency dominated, the scaling of the optimized DeePMD-kit is hindered by the implicit \textit{MPI\_Barrier} in extremely large-scale calculations.  
To alleviate this problem, we reduce the output frequency to every 20 steps, a common practice in the MD community. 
In addition, we replace the \textit{MPI\_Allreduce} with \textit{MPI\_Iallreduce} to further avoid the implicit \textit{MPI\_Barrier}.
% \LL{ the font for MPI operations are inconsistent. Is this intentional?} \WL{changed. }.

%The outputs of the DeePMD-kit are usually global variables such as the total energy, volume, stress, etc.. \textit{MPI\_Allreduce} across all MPI tasks is required to collect these information. The scaling of DeePMD-kit can be hindered by the global collective MPI communication, especially when using the entire Summit supercomputer. 

%\WL{ CONFIRM THE FOLLOWING: In our implementation, the ghost region is communicated via CUDA-aware MPI, which is supported by the IBM Spectrum MPI on Summit. The neighbor list is constructed on the GPU, so that no CPU-GPU memory copy is required when constructing the environment matrix. Note that a global \textit{MPI\_Allreduce} is needed to obtain global properties such as energy and stress whenever output information is needed. Such \textit{MPI\_Allreduce} operations can be replaced by \textit{MPI\_Reduce} to reduce the communication time in large scale simulation, and the frequency of outputting  can be set to once per 10-100 steps. }

\section{Performance measurement}
%\WL{(Note that preference is given to performance actually measured [not projected], based on the entire application [including I/O] and with uniform precision.  Explain in detail if any portion of total runtime was not included in the measurements, if and where different precisions were used, or any attributes listed in Section 3 as “other”). what application(s) was used to measure performance (1 p max) system and environment where performance was measured (1 p max)}

\subsection{Physical Systems}\label{sec:physical_system}
%\WH{I restructure the subsection. please check} \WL{clearer than my version.}
%\WH{The following paragraph may be suoable.}
Among various complex physical systems that have been described by DP, we choose two typical and well-benchmarked systems, one insulating (water) and one metallic (copper), to measure the performance of the optimized DeePMD-kit. 
Water is a notoriously difficult system even for AIMD, due to the delicate balance between weak non-covalent intermolecular interactions,% e.g., the hydrogen bond (HB) network and van der Waals (vdW) dispersion, 
thermal (entropic) effects, as well as nuclear quantum effects~\cite{distasio_jr_individual_2014,chen_ab_2017,ko2019isotope}.
We have shown in Refs.~\cite{zhang2018deep,ko2019isotope} that DeePMD can accurately capture such effects in water.
In combination with extensions of the DP formulation to vectors and tensors, the infra-red~\cite{zhang2020dw} and Raman~\cite{grace2020raman} spectra of water have been properly described.
Copper is a representative simple metal, yet a lot of its properties, such as the surface formation energy and stacking fault energies, can be hardly produced well by EFFs.
In Ref.~\cite{zhang2020dpgen}, using a concurrent learning scheme~\cite{zhang2019active}, we have generated an optimal set of \textit{ab initio} training data and realized a DP model for copper with a uniform accuracy over a large thermodynamic region.

For water and copper, the cut-off radii are 6~\AA\ and 8~\AA\ and the cut-off numbers of neighbors are {144 and 512}, respectively. 
The fitting nets of the models are of size {($240, 240,240$)}, 
%\WH{I propose to change it to $(240,240,240)$}, \WL{changed.}
and the embedding nets are of size {($32,64,128$)}.
%\WH{I propose to change it to $(32,64,128)$}. \WL{changed.}
To test the performance, the MD equations are numerically integrated by the velocity-Verlet scheme for 500 steps (the energy and forces are evaluated for 501 times) at time-steps of 0.5~fs (water) and 1.0~fs (copper). 
The velocities of the atoms are randomly initialized subjected to the Boltzmann distribution at 330~K. 
The neighbor list with a 2~\AA\ buffer region is updated every 50 steps. 
The thermodynamic data, including kinetic energy, potential energy, temperature, pressure, are collected and recorded every 20 time-steps.

For the water system, the strong scaling tests are performed on a system with {4,259,840 molecules (12,779,520 atoms)}. The total number of floating point operations for 500 MD steps of this system is {$151.1$} PFLOPs. 
Weak scaling tests ranging from {42,467,328 to 679,477,248} atoms are performed on up to 4,560 computing nodes on Summit. 
%We notice that compared to the water system, the copper system can %be {$3.5$} times bigger in terms of floating point operations %and GPU memory footprint 
%under the same number of atoms.
We notice that compared with the water system, the copper system, with the same number of atoms, has {$3.5$} times more floating point operations.
%\LZ{Is this better? ``We notice that compared with the water system, the copper system, with the same number of atoms, has {$3.5$} times more floating point operations.'' }\WL{changed.}
The strong scaling tests of the copper system are carried out with a system of {15,925,248} atoms. 
The total number of floating point operations for 500 MD steps of this system is {$588.7$} PFLOPs.
% \LL{When counting the total number of floating-point operations, I think it should be PFLOPs (with a lower case s), please check} . 
% \LZ{I didn't find good examples but I found a lot like xxx PFLOP/s, e.g., https://www.top500.org/lists/2019/06/highs/,
% will leave to Weile for an additional check.}\WL{Lin is right, changed. }
The weak scaling tests are performed on up to 4,560 computing nodes of Summit for systems ranging from {7,962,624 to 127,401,984} atoms. 

Since the baseline DeePMD-kit is restricted by its sequential implementation and can run none of these systems, a fraction of the water system (12,288 atoms/4096 water molecules) is used for comparison with the optimized code on a single GPU in Sec.~\ref{subsec:single_gpu}.

% \begin{itemize}
%     \item For the water systems, the strong scaling tests are performed on a $4,259,840$ molecules ($12,779,520$ atoms) system. The total number of floating point operation for 500 MD steps of this system is $124.83$ PFLOP. Weak scaling tests ranging from $42,467,328$ to $679,477,248$ atoms are performed on up to 4560 computing nodes on Summit. 
%     \item  The strong scaling tests of the copper system are carried out with a system of $15,925,248$ atoms. The total number of floating point operation for 500 MD steps of this system is $835.53$ PFLOP. And weak scaling tests are performed on up to $4560$ computing nodes of Summit for systems ranging from $7,962,624$ to $127,401,984$ atoms. 
% \end{itemize}

\subsection{HPC Platforms and Software Environment }

%The Summit supercomputer, which ranks No. 1 on the top500 list ~\ref{top500_list}, is used for performing all of our numerical tests. 

All the numerical tests are performed on the Summit supercomputer, which consists of 4,608 computing nodes and ranks No.~$2$ on the TOP500 list for a peak performance of 200 PFLOPS~\cite{Top500list}. 
%The 4608 nodes are mounted in 256 racks with 18 computing nodes per rack. 
Each computing node has two identical groups, each group has one POWER 9 CPU socket and 3 NVIDIA V100 GPUs and they are interconnected via NVLink. 
The total computing power for a single node is 43 TFLOPS in double precision (each V100 GPU $7$ TFLOPS and each POWER 9 socket $515$ GFLOPS, thus 7$\times$6$+$2$\times$0.5=43 TFLOPS in total), 86 TFLOPS in single precision, {and 720 TFLOPS in half precision with Tensor Cores (120 TFLOPS per GPU).}
%We remark that 6 V100 GPUs provide 720 TFLOPS half precision computing power with Tensor cores, but they are not used in our tests due to accuracy problem. 
Each computing node has 512 GB host memory and $96$GB ($16$GB per GPU) GPU memory. 
The CPU bandwidth is 135 GB/s (per socket) and GPU bandwidth is 900 GB/s (per GPU). 
The two groups of hardware are connected via X-Bus with a 64 GB/s bandwidth. 
The computing nodes are interconnected with a non-blocking fat-tree using a dual-rail Mellanox EDR InfiniBand interconnect with a total bandwidth of 25 GB/s.

\begin{table}[ht]
\caption{\label{tab:software} Software environment}
\begin{tabular}{ l l } 
 \toprule
 Name & Module used \\
 \midrule
 MPI  & IBM Spectrum MPI 10.3.1.2-20200121 \\
 Host compiler & GCC 4.8.5 \\ 
 GPU compiler & CUDA 10.1.168 \\
 TensorFlow   & IBM-WML-CE 1.6.2-2 (TensorFlow 1.15 included) \\
\bottomrule
\end{tabular}
\end{table}
The software environment is listed in Table ~\ref{tab:software}. In all tests, a single OpenMP thread is used. We use 6 MPI tasks per computing node (3 MPI tasks per socket to fully take advantage of both CPU-GPU affinity and network adapter), and each MPI task is bound to an individual GPU.

\subsection{Measurements}
The total number of floating point operations (FLOPs) 
of the systems is collected via the NVIDIA CUDA NVPROF tool. 
We remark that NVPROF only gathers the FLOPs on the GPU. 
However, in DeePMD-kit, all computationally intensive calculations are performed on the GPU, thus the total FLOPs is reasonable. 
Both double-precision and mixed-precision results are reported in Sec.~\ref{sec:results}. 
The following three criteria are used to measure the performance of the DeePMD-kit. 
%Two measurements, the time-to-solution and sustained performance are measured with wall clock time. The time-to-solution is defined as the the average wall clock time used in calculating a single MD step. 
\begin{itemize}
    \item \textbf{Time-to-solution}, defined as $\frac{ \textsf{MD loop time}} {\textsf{number of MD steps}}$, the average wall clock time used for calculating a single MD step
    %, and calculated by 
    . The ``MD loop time'' includes all the time used in the MD loop (IO included). 
    Setup time, such as the setup of the system and MPI initialization and finalization, is not included. 
    \item \textbf{Peak performance}, defined as $\frac{ \textsf{total FLOPs}} {\textsf{MD loop time}}$. 
    \item \textbf{Sustained performance}, defined as $\frac{ \textsf{total FLOPs}} {\textsf{total wall clock time}}$. The ``total wall clock time'' includes the whole application running time (including IO).%, but MPI initialization and finalization time are excluded. 
\end{itemize}
%Although sustained performance is reported in this paper, MD simulation usually takes hours in practice, thus the overhead (usually a few seconds) is negligible, and sustained performance is nearly the same as the peak performance. 
%\WH{We have a section discussing sustained performance and the information are presented there, I would recommend removing this sentence.}\WL{then I will mark it as ``suoable''}

\section{Performance Results}\label{sec:results}
%\WL{include scalability (weak and strong), time to solution, efficiency (of bottleneck resources), and peak performance (2 pp max)}

% In this section, we evaluate the performance of the optimized DeePMD-kit. 
% First, we compare the optimized DeePMD-kit with the baseline code (sequential implementation) on single GPU using a water system of $12,288$ atoms. 
% Then we measure the performance aspects of the optimized DeePMD-kit on both a $12,779,520$-atom water system and a $15,925,248$-atom copper system, using up to 4560 computing nodes on Summit (strong scaling). 
% Next, we measure the weak scaling of both copper and water systems using from $570$ to $4560$ computing nodes on Summit. 
% Last we show the sustained performance of the DeePMD-kit on a $127,401,984$-atom copper system. \WH{Information in this paragraph has been delivered in 6.1, we may remove it to save space.}\WL{totally agree.}

\subsection{Single GPU}\label{subsec:single_gpu}
%In this part, we show the performance of the individual components, including the customized TensorFlow operators, the standard TensorFlow operators, and the total performance of a single GPU on Summit.
%\LZ{My suggested change: 
In the following, taking the double-precision implementation as an example, we discuss our optimizations on the customized and standard TensorFlow operators in Secs.~\ref{customized_tf_op} and~\ref{standard_tf_op}, respectively.
Then we discuss the implementation of mixed precision and the overall performance in Sec.~\ref{sec:perf-mixed}.
%}
%All the numerical results are measured on a single Summit node. 
 
\subsubsection{Customized TensorFlow operators}  \label{customized_tf_op}

We optimize the customized TensorFlow operators with CUDA customized kernels according to Sec.~\ref{sec:gpu_optimization}. 
In the baseline implementation, the customized TensorFlow operators take about 85\% of the total MD loop time for a water system of 12,288 atoms. 
% A step-by-step optimization of the neighbor lit formatting is shown in Fig.~\ref{fig:environment-performance}~(a), and the wall clock time for each individual customized operator is demonstrated in Fig.~\ref{fig:environment-performance}~(b). 
The performance of the customized operators of the baseline and optimized implementations are compared in Table ~\ref{tab:customized_opt}.
For all the customized TensorFlow operators, an overall speedup of $64.6$ times is achieved. 
Moreover, a total speedup factor of $6.2$ is reached for the ``MD loop time''.% compared to the baseline.

% weile, will add this table, and remove next figure. 
\begin{table}[]
\caption{\label{tab:customized_opt} Performance of optimized customized TensorFlow operators. Baseline customized operators are implemented on CPU.}
\begin{tabular}{ l @{\hspace{3em}} d{4.2} d{4.2} d{3.0} } 
\toprule
Operators & \tabcenter{Baseline[ms]} & \tabcenter{Optimized[ms]} & \tabcenter{Speedup}\\
\midrule
Environment  & 302.54 & 2.32 & 130\\
ProdViral    &  51.06 & 1.34 &  38\\
ProdForce    &  41.29 & 2.41 &  17\\
\bottomrule
\end{tabular}
\end{table}

% \begin{figure}[]
%   \begin{center}
%   {\includegraphics[width=0.25\textwidth]{environment-1.png}} 
%   \end{center}  
%   \caption{ 
%     Wall clock time (in milliseconds) of different customized TensorFlow operators in one MD step for a water system of 12,288 atoms. 
%     %\WL{(a) start from 93.55 for the CPU}\LuDh{changed from Intel Xeon Gold 6132 CPU to IBM POWER 9 CPU} \WL{thanks}
%   }
%   \label{fig:environment-performance}
% \end{figure}

\subsubsection{Standard TensorFlow operators}\label{standard_tf_op}

Some of the standard TensorFlow operators are re-implemented and optimized according to Sec.~\ref{sec:nn}.
For the water system of 12,288 atoms, MATMUL+SUM, CONCAT+SUM, and TANH+TANHGrad in the baseline implementation are accelerated by 1.3, 1.7, and 1.6 times with GEMM, GEMM, and merged TANH
%in the optimized implementation
, respectively.
% The wall clock time and the corresponding speedup factor for calculating a water system of $12,288$ atoms 
% %using both baseline and optimized DeePMD-kit 
% are shown in Table. ~\ref{tab:nn_opt}.\LZ{Table??}
{The baseline implementation calls standard TensorFlow operators, which are already highly efficient on GPUs,}
% \WH{@WL, please check if this is OK} \WL{yes, as long as you keep ``highly efficient'' LOL}
% \WL{We remark that the speedup is achieved compared to the TensorFlow, a highly efficient GPU library on high-performance supercomputers. } 
yet an extra $1.21$ 
% \LuDh{loop time: 1.24893 vs 0.978345}
times of speedup is achieved for the ``MD loop time'' compared with Sec.~\ref{customized_tf_op}.
% \WH{The numbers should be updated!}
%An overall of $xx$ times of speedup is reached for the ``MD loop time'' compared to the previous state-of-art after this optimization.

%\WL{comment on the Half precision............}
%\begin{table}[ht]
%\caption{\label{tab:gemm} Matrix Multiplication Performance}
%\small
%\begin{tabular}{ l r r r} 
% \toprule
% GEMM & NN & NT & TN \\
% \midrule
% \%of Peak   & 52.86 &  64.08 & 71.17 \\
%\bottomrule
%\end{tabular}
%\end{table}

\subsubsection{Mixed precision}\label{sec:perf-mixed}

\begin{table}[]
    \caption{\label{tab:test-err}Test errors for the water system from models with different precision.}
    \centering
    \begin{tabular}{lcc}
    \toprule
    Precision & Error in energy [eV/molecule] & Error in force [eV/\AA] \\ \midrule
       Double  & $1.2\times10^{-3}$ & $3.7\times 10^{-2}$\\
       MIX-32  & $1.2\times10^{-3}$ & $3.7\times 10^{-2}$\\
       MIX-16  & $3.6\times10^{-3}$ &
       $3.8\times 10^{-2}$\\
    \bottomrule 
    \end{tabular}
    \label{tab:my_label}
\end{table}

% We compare the prediction of the mixed precision models with that of the double precision model for a well equilibrated water system with 512 molecules.
% Deviations of $5.2\times10^{-6}$ and $2.5\times10^{-3}$~eV (normalized by number of molecules) in energy and $4.3\times 10^{-6}$ and $5.7\times 10^{-3}$~eV/\AA\ in forces are observed \wl{for the MIX-32 and MIX-16 codes, respectively}. 
% The deviation of the MIX-32 is essentially smaller than the prediction error of the double precision model compared with \textit{ab initio} data.

\begin{figure}
  \begin{center}
  {\includegraphics[width=0.47\textwidth]{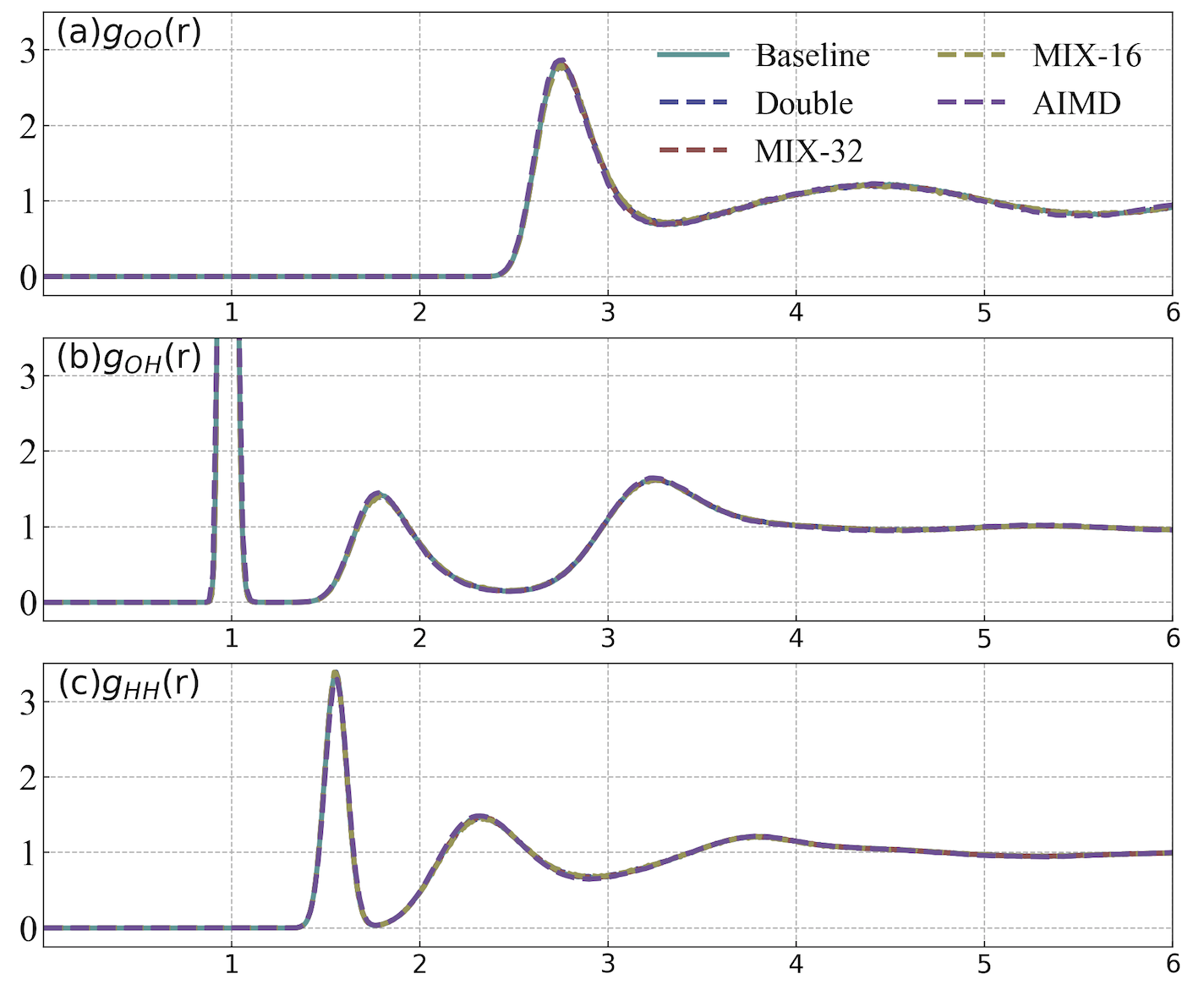}}
  \end{center}
  \caption{ 
    Radial distribution functions $g_{OO}(r)$, $g_{OH}(r)$, and $g_{HH}(r)$ of liquid water at ambient conditions, calculated by AIMD and four DeePMD-kit implementations: baseline, optimized double, MIX-32, and MIX-16.
    %\LuDh{Loop time of 6836.88 on 1 procs for 60000 steps with 12288 atoms. Performance: 0.379 ns/day, 63.304 hours/ns, 8.776 timesteps/s.} 
  }
  \label{fig:precisions}
\end{figure}

The accuracy of the mixed precision models is investigated by comparing the energy and forces computed from DeePMD-kit with those from AIMD predictions. We take water as an example and the test data set is composed of 100 water configurations of 64 molecules.
As shown in Table~\ref{tab:test-err},
the MIX-32 scheme is as accurate as the double precision. 
The accuracy of MIX-16 is slightly worse than that of the double precision model, but is usually enough for an accurate prediction of physical observables.
To further check the accuracy, we calculate the radial distribution function (RDF), the normalized probability of finding a neighboring atom at the spherically averaged distance $r$.
The oxygen-oxygen ($g_{OO}(r)$), oxygen-hydrogen ($g_{OH}(r)$), and hydrogen-hydrogen ($g_{HH}(r)$) RDFs are typically utilized to characterize the structures of water~\cite{zhang2018deep}.
%, and they are compared 
As shown in Fig.~\ref{fig:precisions}, the RDFs computed from the mixed-precision implementations (MIX-32 and MIX-16) agree perfectly with those from the double-precision implementation and those from the AIMD calculation.
Therefore, we conclude that the mix-precision methods do not lead to loss of accuracy for predicting physical observables.
 
For the water system, compared with the double-precision version, the {MIX-32} code is about $1.7$ times faster and saves half of the GPU memory cost, and the MIX-16 code is $2.6$ times faster and saves ~$75\%$ of the GPU memory cost. 
% \WH{@WL in water system, the numbers at weak scaling limit are 1.7 and 2.6; in the copper system, the numbers are 1.8 and 3.0. Please recheck what we want to present.}
%, compared to the double-precision version. 
Together with the speedups from Secs.~\ref{standard_tf_op} and ~\ref{customized_tf_op}, it is concluded that the optimized DeePMD-kit with double precision is around {7.5}
% $11.3$
times faster than the baseline code,
% (Table~\ref{tab:soa}),
{and the speedup factor increases to {12.7 and $19.5$} when the {MIX-32 and MIX-16 codes are used.}
%mixed precision is used.
%If the mixed precision is used then an overall 11.3 times of speedup is achieved.
}

\begin{figure}
    \begin{center}
    {\includegraphics[width=0.45\textwidth]{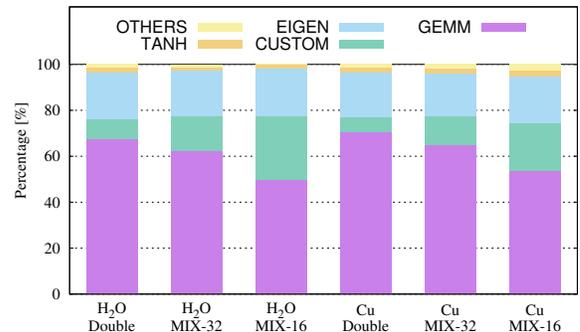}} 
    \end{center}
    \caption{
        Percent stacked bar chart of different TensorFlow operators in terms of the GPU computational time for $24,576$ and $4,860$ atoms for water and copper, respectively.
        %\LL{ add the number of atoms for H2O and Cu here in the caption} \WL{added.}
        }
    \label{fig:proportions}
\end{figure}

Finally, Fig. ~\ref{fig:proportions} shows the percentage of time spent by different TensorFlow operators in the total GPU execution time.  
We notice that the contribution from the GEMM operator is more important in the copper system (double: $71\%$, MIX-32: $65\%$ MIX-16: $54\%$) than that in the water system (double: $67\%$, {MIX-32}: $62\%$, {MIX-16: $50\%$}).
This is mainly attributed to the fact that the FLOPs of the copper system is $3.5$ times bigger than that of the water due to the larger number of neighbors per atom, as discussed in Sec.~\ref{sec:physical_system}.
%\LuDh{I think the difference between water and copper system is mainly caused by the four separate embedding-net of water system. We do not have slicing and sorting operator in both water and copper system. We need talk here. "the FLOPs of the copper system is $3.5$ times bigger than that of the water" is correct only when the two systems have the same atoms number}\WL{Thanks. }
We remark that the GEMM operator in DeePMD-kit is still memory-bound due to the the small network size (the dimensions of the three embedding network layers are {32, 64, and 128}). 
Profiling on the water system shows that the average computational efficiency of the GEMM operations is $66.4\%$, $62.3$ and $19.3\%$ for double, MIX-32, and MIX-16 versions, respectively.  
The corresponding bandwidth utilization is $88.9\%$, $88\%$ and $87.6\%$ of the hardware limit, respectively. {As the network size grows, 
% the FLOPS of the MIX-16 will increase. 
the bandwidth limitation will be alleviated.
A detailed discussion will be presented in Sec.~\ref{sec:network_size}}.

\subsection{Scaling}\label{sec:scaling}

We discuss the scaling behaviors of the optimized DeePMD-kit on the Summit supercomputer for large-scale simulations. 
%The system sizes are inaccessible with the baseline implementation. Specifically, 
The system sizes, ranging from 8 to 679 million of atoms, 
are inaccessible with the baseline implementation, and are more than two orders of magnitude larger than other state-of-the-art MD schemes with \textit{ab initio} accuracy. 

\subsubsection{Strong Scaling}
\begin{figure}
  \begin{center}
    \includegraphics[width=0.5\textwidth]{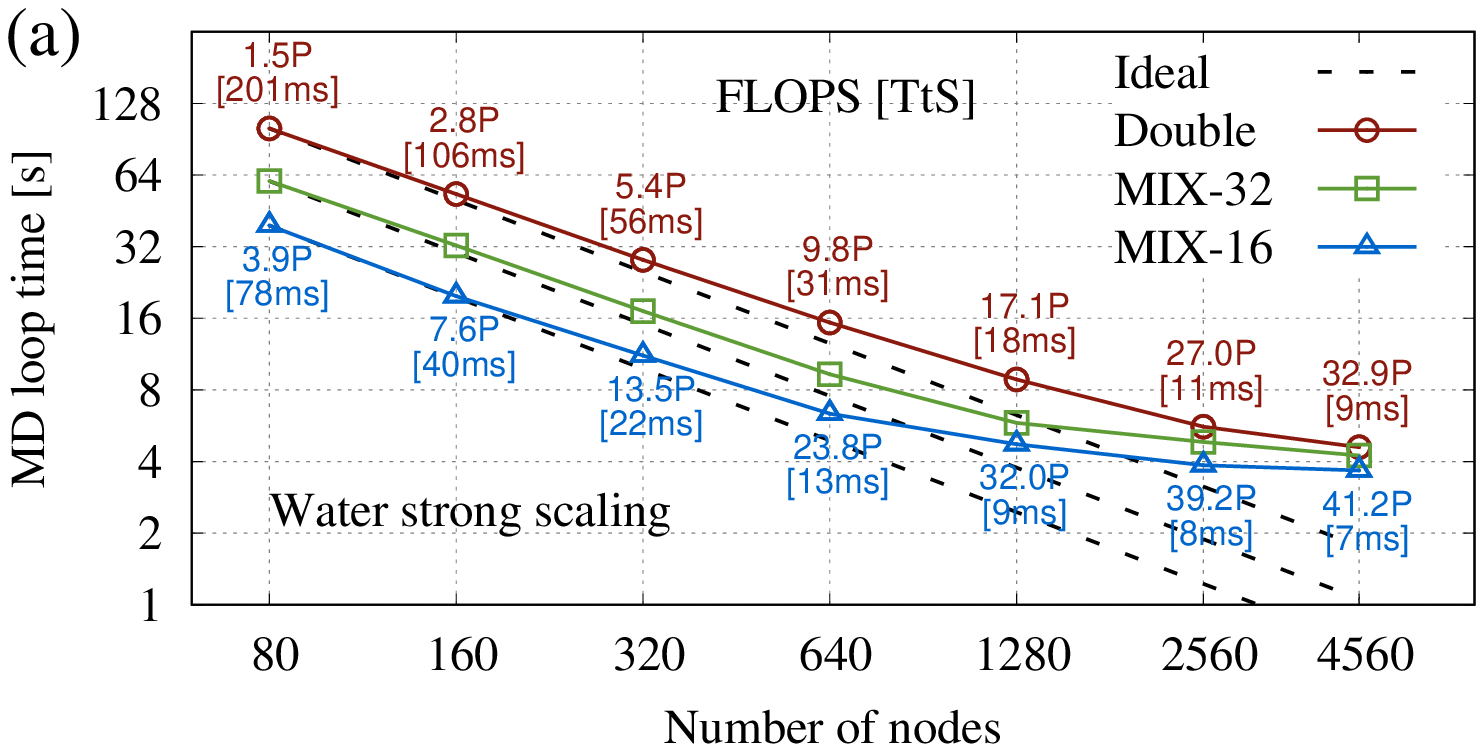} \\
    \includegraphics[width=0.5\textwidth]{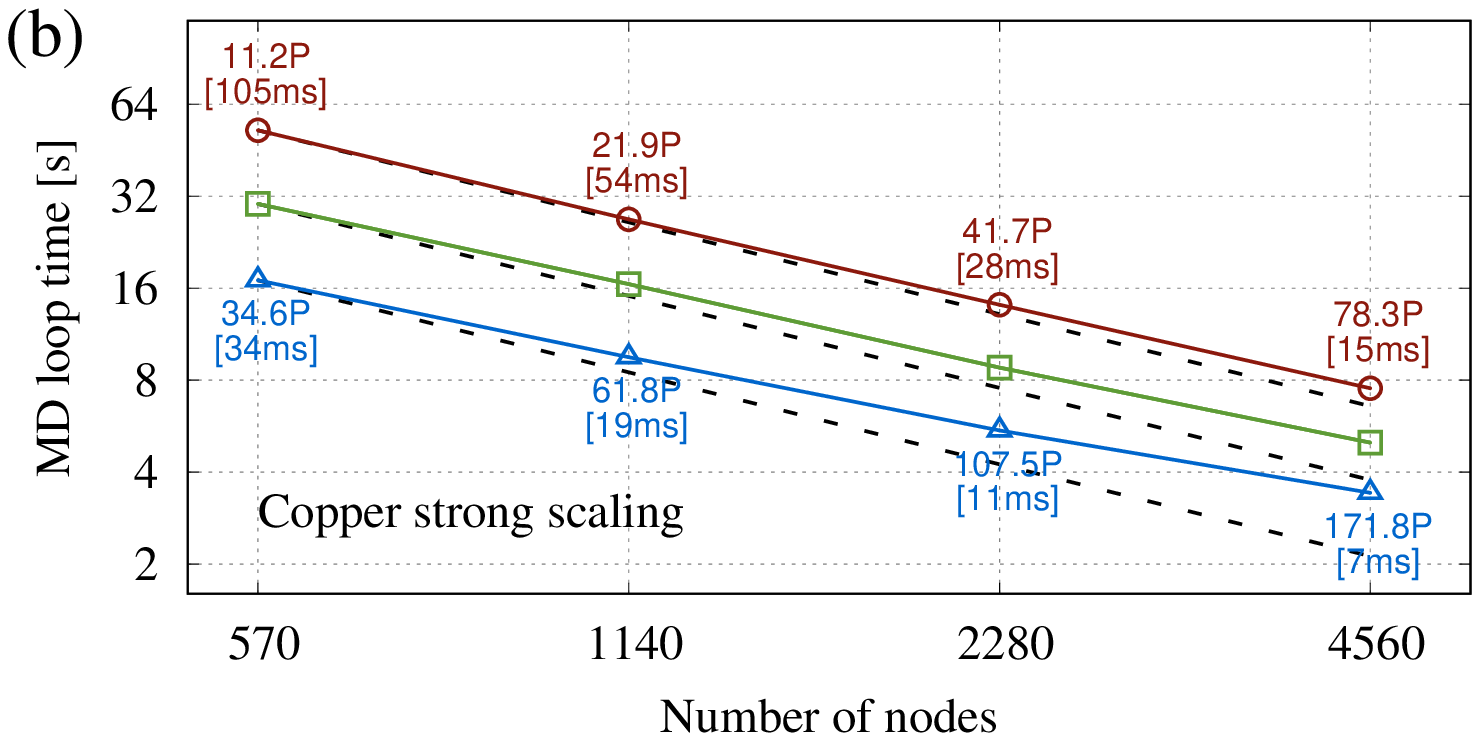}
  \end{center}
  \caption{Strong scaling: (a) the water system of 12,779,520 atoms. (b) the copper system of 15,925,248 atoms. The FLOPs of the copper system is {3.9 times} compared  to that of the water system. %\wl{and memory usage}. 
  The MD loop time is measured using the wall clock time of 500 MD steps (energy and forces are computed 501 times). The corresponding peak performance in PFLOPS and the time-to-solution (TtS) in millisecond per MD step are presented.}% along with the data points. }
  \label{fig:strong_scaling}
\end{figure}

In Fig.~\ref{fig:strong_scaling}, we measure the scalability of the optimized DeePMD-kit with the ``MD loop time'' of 500 MD steps ranging from 80 to 4,560 computing nodes. 
The testing systems include a copper system of 15,925,248 atoms and a water system of 12,779,520 atoms. 
%Compared with the water system, the copper system is about $7$ times larger in terms of both memory usage and total number of floating point operations.
%\LZ{has been mentioned in the figure caption}

For the copper system, the optimized DeePMD-kit scales well to the entire Summit supercomputer. 
By setting the performance with 570 computing nodes as baseline, the parallel efficiency is {$87.3\%$, $71.9\%$, and $61.9\%$} when scaling to 4,560 computing nodes on Summit, \textbf{reaching peak performance of {78.3, 112.3, and 171.8} PFLOPS for the double, MIX-32, and MIX-16 versions of the code, respectively}. 
The time-to-solution of a single MD step for this particular system is {$7$} milliseconds using the {MIX-16} version of optimized DeePMD-kit, making it possible to finish nanosecond simulation within {$2$} hours (time-step 1.0~fs) 
with \textit{ab initio} accuracy. 
%\LZ{Time step has been reported in 6.1?}

For the water system, the optimized DeePMD-kit scales almost perfectly up to 640 computing nodes, and continues to scale up to the entire Summit supercomputer. 
Compare to the baseline of 80 computing nodes, the parallel efficiency of the optimized DeePMD-kit is {$81.7\%$(double), $81\%$(MIX-32) and $77\%$(MIX-16)} when scaling to 640 computing nodes, and decreases to {$38.3\%$(double), $24.9\%$(MIX-32) and $18.7\%$(MIX-16)} when using 4,560 computing nodes. The decrease of the parallel efficiency is mainly due to the scaling of the data size per GPU. As shown in Table ~\ref{tab:data}, the percentage of peak performance goes down dramatically when the number of atoms per GPU is less than 3,000, especially for the MIX-16 code.
%\WH{shall we say ~1000?  usually 50\% is though to be an acceptable efficiency, the water system decreases to 50\% at 819 atoms/GPU} \WL{yes, changed.}
However, {we remark that all double and mixed-precision versions of DeePMD-kit scale up to 4,560 computing nodes with $459$ atoms per GPUs despite the small data size. \textbf{The time-to-solution of a single MD step for this system with double-precision is 9 milliseconds, making it possible to finish nanosecond simulation in 5 hours
(time-step is 0.5 fs).
%\LZ{Time step has been reported in 6.1?}
}}

\begin{table}[ht]
\caption{\label{tab:data} Average number of atoms (per GPU), average ghost region size (per GPU), and double precision FLOPS for the {12,779,520} atoms water system.}
{\scriptsize
\begin{tabular}{ l l l l l l l l} 
 \toprule
 \#Nodes & 80 & 160 & 320 & 640 &1280 & 2560 & 4560\\
 \#GPUs & 480 & 960 & 1920 & 3840 &7680 & 15360 & 27360\\
 \midrule
 \#atoms  & 26624  &13312  &6656  &3328  &1664  &832  &467 \\
 \#ghosts  & 25275  &17014  &11408  &7839   &5553  &3930   &3037 \\
 \midrule
 MD time & 100.4  &53.2  &28.1  &15.4  &8.8  &5.6  &4.6\\
 Efficiency & 1.00  &0.94  &0.89  &0.82  &0.71  &0.56  &0.38\\
 PFLOPS   & 1.51  &2.84  &5.37  &9.84  &17.09  &26.98  &32.90 \\
 \%of Peak   & 42.90  &40.45  &38.26  &35.07  &30.44  &24.03  &16.45 \\
\bottomrule
\end{tabular}}
\end{table}

\subsubsection{Weak scaling}

\begin{figure}
  \begin{center}
    \includegraphics[width=0.5\textwidth]{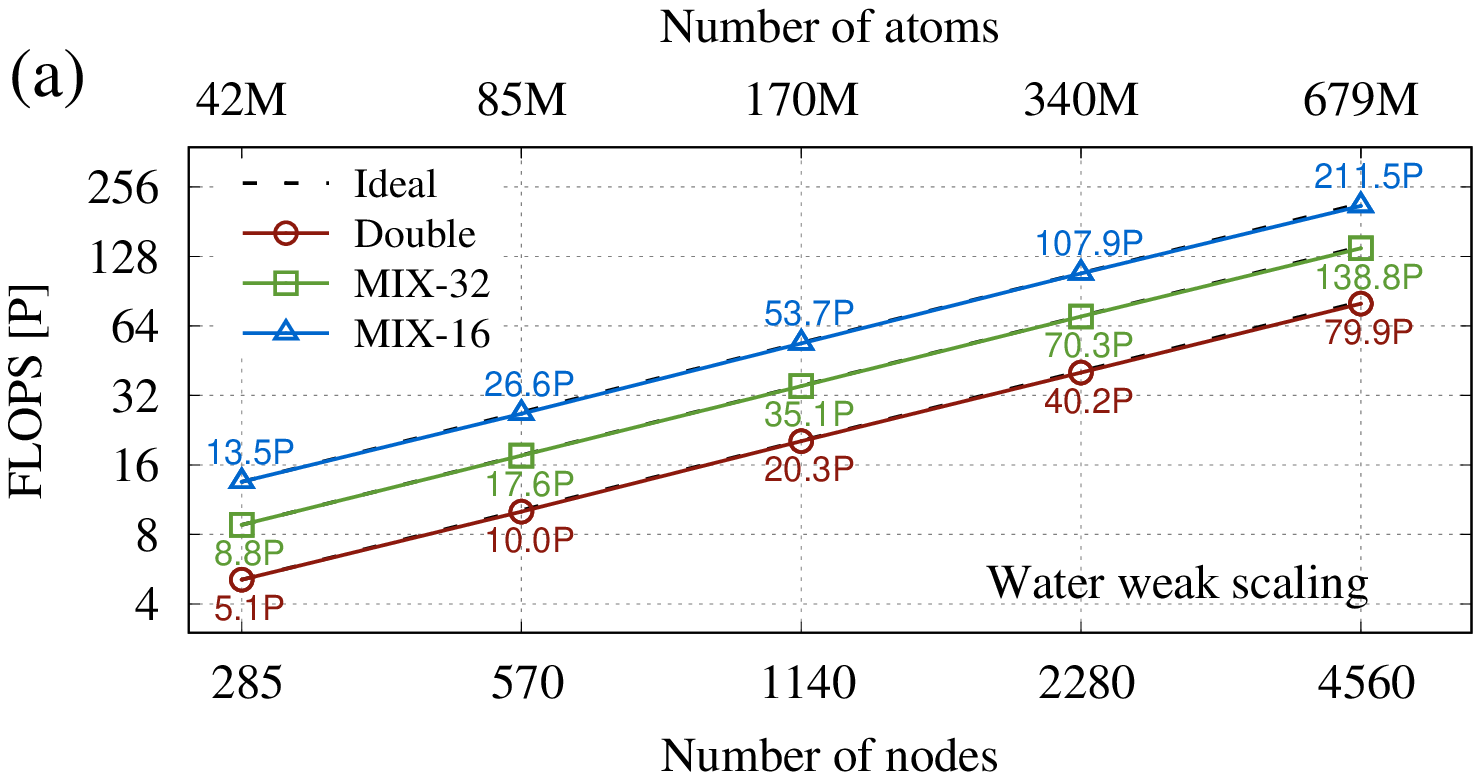} \\\vskip .5cm
    \includegraphics[width=0.5\textwidth]{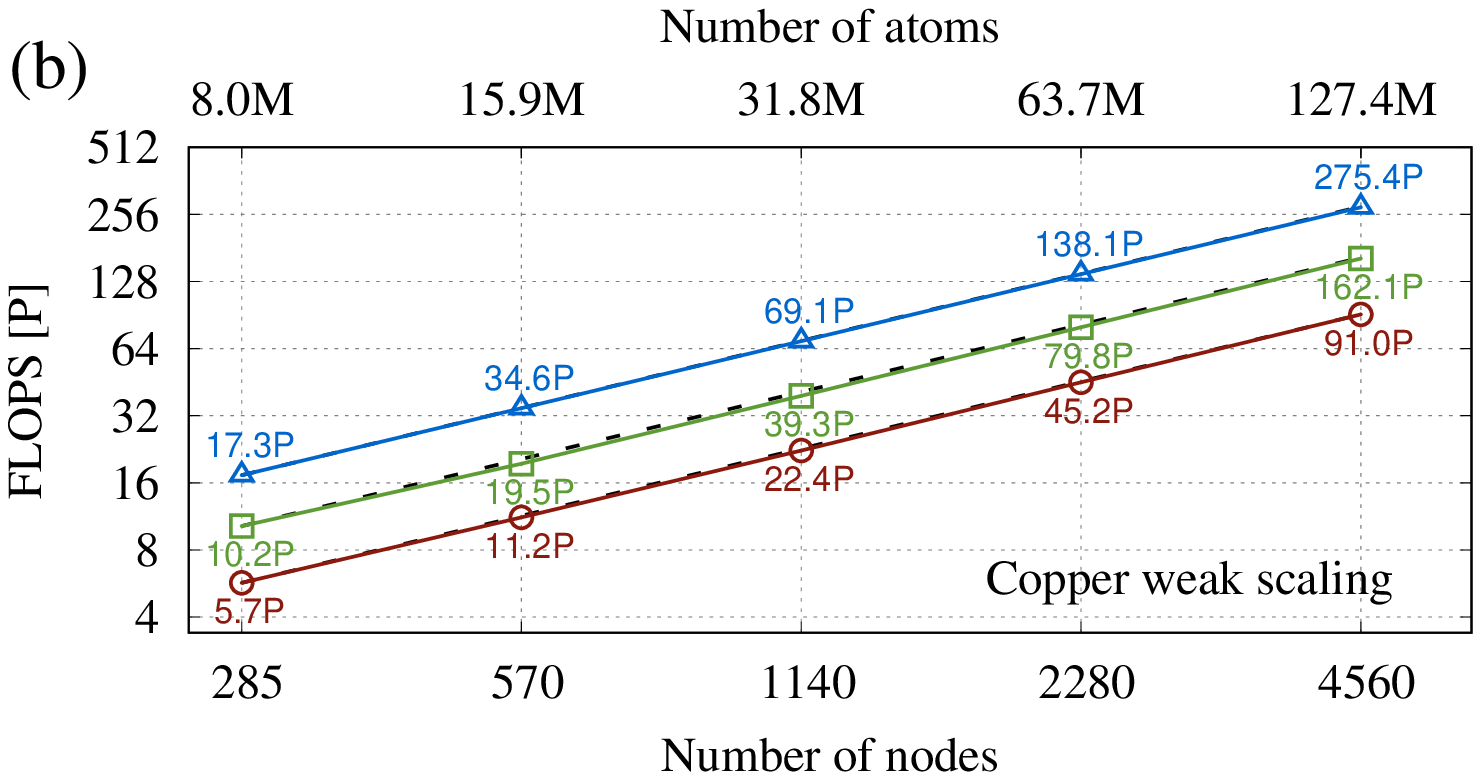}
  \end{center}
  \caption{Weak scaling: 
  (a) the water system. Number of atom ranges from 42,467,328 to 679,477,248.
  (b) the copper system. Number of atoms ranges from 7,962,624 to 127,401,984. }
  \label{fig:weak}
\end{figure}

The weak scaling of the optimized DeePMD-kit is measured in terms of the FLOPS of 500 MD steps for both water and copper (Fig.~\ref{fig:weak}). 
%\WH{In Fig.~\ref{fig:weak} we do not present the "MD loop time"} \WL{changed.}
Both systems show perfect scaling with respect to the number of nodes (GPUs) used. 
The {MIX-32 and MIX-16 versions} are about {$1.7$/1.8 and $2.6$/3.0} times faster compared to the double-precision code for the water/copper system, respectively. 
% \WH{The acceleration by MIX is different in water and copper systems. please recheck the numbers.}
For water and copper, the largest system sizes simulated in these tests are {679 and 127 million atoms}, respectively, which are more than {three} orders of magnitude larger compared to the state-of-the-art MD with \textit{ab initio} accuracy. 
For the copper system, \textbf{the peak performance achieved is 91 PFLOPS (45.5\% of the peak%, 98.7\% parallel efficiency
) in double precision, and 162/275 PFLOPS in MIX-32/MIX-16 precision.}
%The time-to-solution is {$8.1\times 10^{-10}$} second/step/atom in double precision, and {$4.6\times 10^{-10}$} 
%\WH{recheck this number! by which we do not have 1.7 times speed up}
%\WL{all the highlighted numbers are not final, I have re-calculated them. Now they are correct. } 
%The time-to-solution is {$8.1\times 10^{-10}$} second/step/atom in double precision, and {$4.6\times 10^{-10}$} and $2.7\times 10^{-10}$ second/step/atom in MIX-32 and MIX-16 precision, respectively, which means that one nanosecond MD simulation of the 127M-atom system with \textit{ab initio} accuracy can be finished in {29, 16 and 9.5} hours (1.0~fs time-step), respectively. 
The time-to-solution is {$8.1/4.6/2.7\times 10^{-10}$} second/step/atom in double/MIX-32/MIX-16 precision, which means that one nanosecond MD simulation of the 127M-atom system with \textit{ab initio} accuracy can be finished in {29/16/9.5} hours.
For the water system, \textbf{the peak performance is {79.9 PFLOPS (40\%} of the peak) in double precision, and {138.8/211.5} PFLOPS in {MIX-32/MIX-16} precision.} 
%\WL{Note: MIX-16 is 2.64x faster than double precision for water, but 3x faster for copper. }
%The optimized code reaches a time-to-solution of {$3.0\times 10^{-10}$ second/step/atom} in double precision and {$1.7\times 10^{-10}$}, {$1.1\times 10^{-10}$} second/step/atom in {MIX-32 and MIX-16} precision. Thus one nanosecond MD simulation of the {679M-atom} water system with \textit{ab initio} accuracy can be finished in {112, 64 and 42} hours (0.5~fs time-step) in double, {MIX-32 and MIX-16} precision, respectively. 
The optimized code reaches a time-to-solution of {$3.0/1.7/1.1\times 10^{-10}$ second/step/atom} in double/MIX-32/MIX-16 precision, so that one nanosecond MD simulation of the {679M-atom} water system with \textit{ab initio} accuracy can be finished in {112/64/42} hours.
% \textbf{The peak performance achieved is 72.6 PFLOPS (36\% of peak) or 91.2 PFLOPS (45.5\% of peak) with double precision for the water or copper system, respectively. And the corresponding mixed-precision peak is 105.4 PFLOPS or 162.4 PFLOPS, respectively. }  
% The time-to-solution is $110$ or $76$ milliseconds per step (61 or $42$ hours per nanosecond, time-step 0.5 fs) for the water system with double or mixed precision, and the corresponding time-to-solution for copper is $83$ or $52$ milliseconds per step (23 or $14$ hours per nanosecond, time-step 1.0 fs), respectively. 
% \LL{Need time to solution per nanosecond to be consistent with what we highlighted earlier in the paper.} \WL{added}
%{Since the system size is limited by the capacity of the GPU memory, the MIX-32 and MIX-16 codes can accommodate physical systems of $1.35$ and $2.7$ billion atoms with 4,560 computing nodes.}
%We remark that the perfect linear scaling of both systems implies that the optimized DeePMD-kit is able to calculate even bigger physical systems on future exascale supercomputers with no intrinsic obstacles. 
{We remark that the computationally feasible system sizes for MIX-32 and MIX-16 codes on the 4,560 computing nodes of Summit can keep increasing to 
%accommodate physical systems of 
$1.35$ and $2.7$ billion atoms, respectively, and will be ultimately limited by the capacity of the GPU memory.
Moreover, the perfect weak scaling of both systems implies that the optimized DeePMD-kit is able to calculate even bigger physical systems on future exascale supercomputers with no intrinsic obstacles.}

\subsection{Sustained performance}
The MD loop time of the optimized DeePMD-kit has been measured and discussed in detail in Secs.~\ref{subsec:single_gpu} and \ref{sec:scaling}. 
By subtracting the MD loop time from the total wall clock time, we define the ``setup time'', which mainly includes the initialization of the atomic structure and the loading of the DP model data.
In the baseline implementation, the atomic structure is constructed on a single MPI task and then distributed via MPI communication, and the model data is read in from the hard-drive by all the MPI tasks. 
The corresponding setup time can be a few minutes, though they are  performed only once. 
For example, the setup time for the copper system of {$127,401,984$} atoms is more than {$263$} seconds on 4,560 computing nodes on Summit.  
 
To reduce these overheads, we build the atomic structure with all the MPI tasks without communication, and the model data is also staged by first reading in with a single MPI rank, and then broadcasting across all MPI tasks.  
By these optimizations, the setup time is reduced to less than $5$ seconds for all tests. 
\textbf{The sustained performance of the DeePMD-kit reaches {90.3}~PFLOPS (45\% of the peak) in double precision when running the {127,401,984} atoms copper system for 5,000 MD steps (5 ps). }
%The time-to-solution is {$8.1\times 10^{-10}$ second/step/atom, equivalent to 29} hours for a one-nanosecond simulation. 
%\LZ{Do we need to report time-to-solution here?}

\subsection{Network size}\label{sec:network_size}

\begin{figure}
  \begin{center}
    \includegraphics[width=0.5\textwidth]{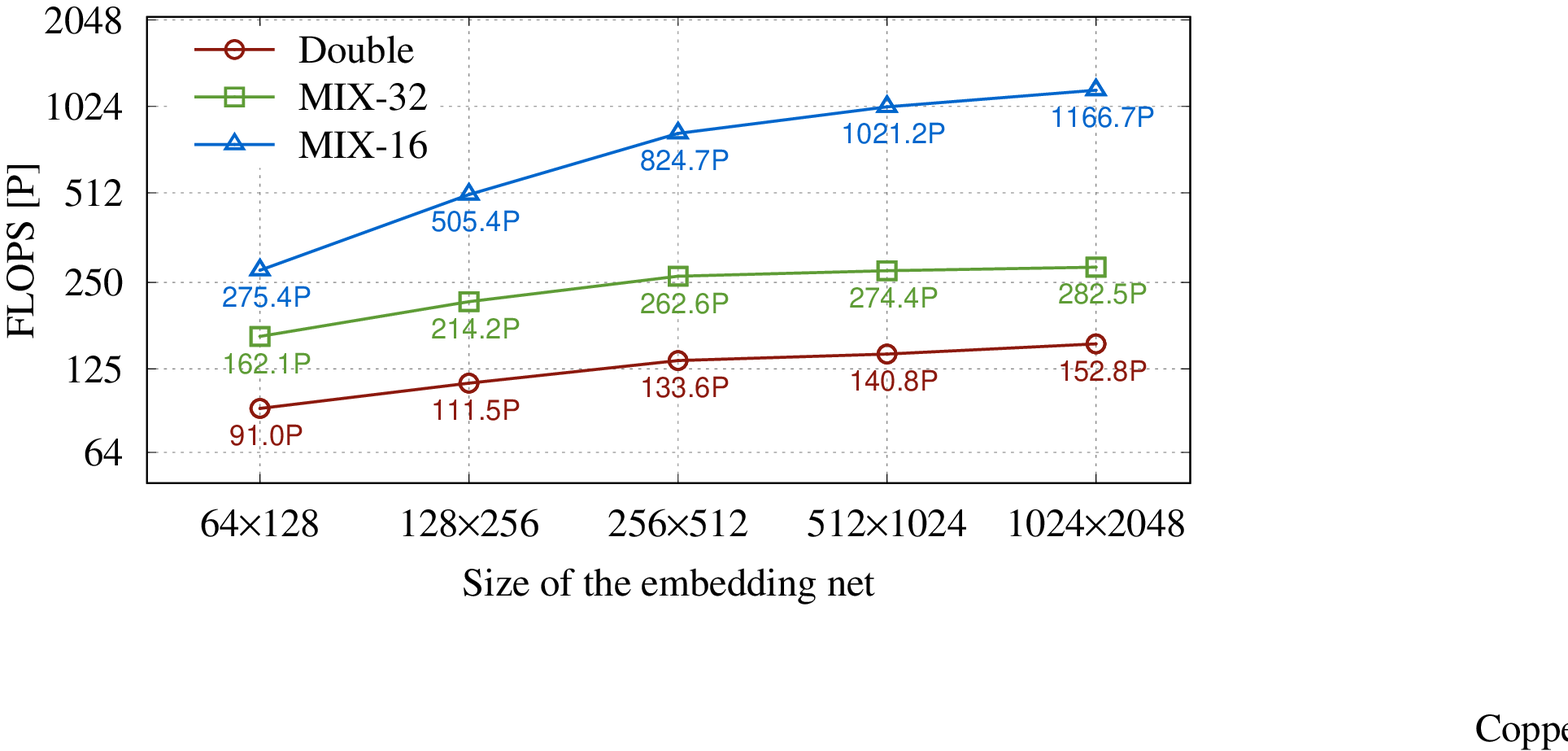} 
  \end{center}
  \caption{The FLOPS achieved with respect of the size of the embedding net. The FLOPS is measured with the copper system %of 127,401,984 atoms \WL{#atom changes with network size. }
  on 4,560 nodes. The size of the embedding net is characterized by the size of the network parameter matrix ${W}$ of the largest hidden layer. The embedding net investigated in this work is 64$\times$128.}
  \label{fig:network_size}
\end{figure}

The performance of the DeePMD-kit with respect to the matrix size of the last layer of the embedding net is shown in Fig.~\ref{fig:network_size}. While the FLOPS curves of both double and MIX-32 versions flatten after the matrix size reaching $256\times512$, that of the MIX-16 version keeps increasing and reaches $1.17$ EFLOPS when the matrix is of size $1024\times2048$. This is mainly because half precision arithmetic is only efficient when the matrix size is bigger than $2048\times2048$~\cite{markidis2018nvidia}. 
Although larger FLOPS comes with bigger networks, we notice that it is enough to achieve the \textit{ab initio} accuracy with matrix size $64\times128$, 
and the accuracy improvement by using larger embedding nets is negligible.
% and provides better time-to-solution for both water and copper. 
% \WH{Here we want to say 64x128 net is enough from the accuracy perspective}
Therefore, in this paper, we report the performance of the DeePMD-kit based on the matrix size $64\times128$. In this regime, the performance is mainly dominated by the GPU memory bandwidth, as discussed in section ~\ref{subsec:single_gpu}. 
As the size of the embedding net grows to $1024\times2048$, the GEMM operations takes more than $85\%$ of the GPU computational time in all versions of DeePMD-kit. 
The computational efficiencies of both double precision and MIX-32 achieve about $90\%$, though that of the half precision only reaches $42\%$,
which indicates that the performance of the double and MIX-32 are compute-bound, and the MIX-16 version is still memory-bound.
Such performance behavior can be understood by the FLOP/Byte ratio of the V100 GPU, and will be discussed in section ~\ref{sec:8-2}.  
We remark that larger network size may be needed to achieve better accuracy in more complicated physical systems than pure water and copper.
%investigated in this work.
In those cases, the MIX-16 scheme is even more favorable in terms of efficiency.
% \LL{ "reach peak performance" seems ambiguous.} 
% \WL{changed to 'FLOPS curve flattens ', better, or not?}
%\WH{We may want to provide memory bandwidth utilization and computational efficiency data for larger network, showing that the bandwidth limitation will be alleviated by larger net thus higher peak performance is achieved.}
%\WL{yes, good suggestion. Will do that. I will not be surprised if the problem is still memory-bound after using larger network size. The reason for the performance increase mainly comes from having larger Matrix size(larger tile size), see arxiv: 1803.04014. }

\section{Implications}
%\WL{implications for future systems and applications (1 p max)}

%\LZ{Prof. E once wrote a few sentences suitable for here. I modified a bit. Not sure we should replace the following paragraph with it or not:
This work provides a vivid demonstration of what can be achieved by integrating physics-based modeling and simulation, machine learning, and efficient implementation on the next-generation computational platform. 
It opens up a host of new exciting possibilities in applications to material science, chemistry, and biology, as introduced in Sec.~\ref{sec:8-1}.
It also poses new challenges to the next-generation supercomputer for a better integration of machine learning and physical modeling, as detailed in Sec.~\ref{sec:8-2}. 
We believe that this work may represent a turning point in the history of high-performance computing, 
and it will have profound implications not only in the field of molecular simulation, but also in other areas of scientific computing.
% and it will have profound implications not only in the field of molecular simulation, but also in almost all areas of scientific computing.
%}

%This work opens a new era in large-scale molecular dynamics simulation with \textit{ab initio} accuracy.
%This unprecedented power results from integrating  physics-based modeling and simulation, machine learning, and efficient implementation on the largest computational platform.
%It makes possible direct studies of various problems, as we introduce in Sec.~\ref{sec:8-1}, and it also poses new challenges to the next-generation supercomputer for a better integration of machine learning and physical modeling, as detailed in Sec.~\ref{sec:8-2}.

\subsection{Applications of Optimized DeePMD-kit\label{sec:8-1}}
\begin{figure}
  \begin{center}
  {\includegraphics[width=0.47\textwidth]{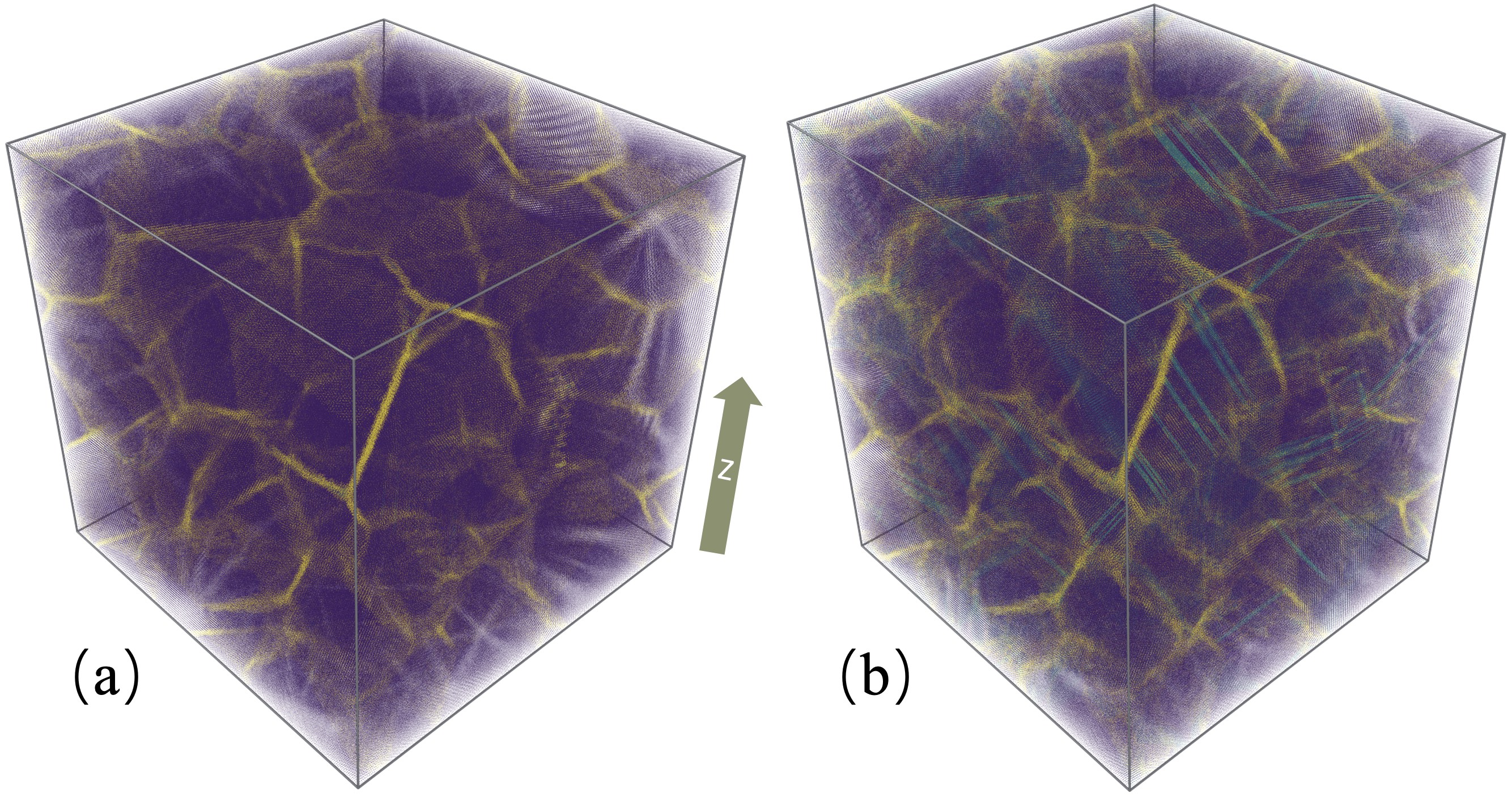}}
  \end{center}
  \caption{ 
    (a) A 10,401,218-atom nanocrystalline copper consisting of 64 randomly oriented crystals with 15-nm averaged grain diameter.
    (b) The nanocrystalline copper after 10\% tensile deformation along the $z$ axis. Purple, yellow, and cyan denote the atoms in the grains, atoms in the grain boundaries, and atoms in the stacking faults.}
  \label{fig:copper}
\end{figure}
The strength and hardness of metals can be enhanced by refining their grains,
and MD can be of great help to provide microscopic insights into the underlying mechanism~\cite{98N-Schiotz,03S-Schiotz}.
Typically, a nanocrystalline structure of metal consists of tens to hundreds of millions of atoms~\cite{98N-Schiotz,03S-Schiotz}, which is far beyond the capability of \textit{ab initio} methods. 
Therefore, previous simulation of nanocrystalline metals can only be driven by EFFs with limited accuracy. 
Taking copper as an example, EFFs are able to yield the strain-stress curves of nanocrystalline, from which the movements of dislocations and grain boundaries can be analyzed to elucidate the origins of strength in nanocrystalline. 
However, the biggest problem of EFFs  is the lack of accuracy for certain properties, e.g., surface formation energies and stacking fault energies.
%as compared to the \textit{ab initio} models~\cite{zhang2020dpgen}; these properties are important for studying mechanical properties of copper, including its nanocrystalline structure. 
The accuracy problem is largely resolved by the DP model used in this work.
We refer to Ref.~\cite{zhang2020dpgen} for extensive benchmark.

We show in Fig.~\ref{fig:copper} the tensile deformation of a 10,401,218-atom  nanocrystalline copper by MD simulations. 
The initial cell size is set to $50\times50\times50$ nm$^3$.
We run 50,000 steps with a time-step of 0.5 fs.
The first 10,000 steps are used for annealing at 300 K while the remaining 40,000 steps follow a strain rate of 5$\times$10$^8$ s$^{-1}$. 
In total, the nanocrystalline copper is deformed by 10\%.
We adopt the common neighbor analysis scheme~\cite{PhysRevLett.60.2295,PhysRevE.47.3975} to analyze the structure of nanocrystalline copper. 
As shown in Fig.~\ref{fig:copper}, the atoms in the grains have a face-centerd cubic (fcc) local structure, which is the ground-state structure of copper. 
After the deformation, stacking faults of copper are identified by monitoring the formation of hexagonal close-packed (hcp) structures. 
%Additionally, the atoms in the grain boundaries are labeled as an ``unknown'' structure. 
This example demonstrates the dynamical tensile deformation process of a nanocrystalline copper system.
We leave detailed analyses to a future paper that is dedicated to the physics of this process.

Applications enabled by the multi-GPU implementation of the DeePMD-kit code can go far beyond copper and water systems reported here, and can span a wide spectrum of complex materials and molecules.
This first stems from the wide applicability of the DP method to problems in different fields.
Being a general model based on both machine learning and physics, DP inherits the accuracy from first-principles methods and puts on an equal footing the description of atomic interaction in the cases of bio-molecules, insulators, metals, and semi-metals, etc.
This ability of DP is further boosted by this work, which takes advantage of the state-of-the-art supercomputers, and makes simulation of hundreds of millions of atoms with \textit{ab initio}  accuracy a routine procedure.
In the short term, this will directly benefit the study of many problems of practical interests, such as complex chemical reactions~\cite{nakano2007divide,li2015revealing}, electrochemical cells~\cite{jorn2013atomistic}, nanocrystalline materials~\cite{lund2004tension,98N-Schiotz,03S-Schiotz}, irradiation damages~\cite{gao2000atomic}, and dynamic fracture and crack propagation~\cite{vashishta1996crack,vashishta1999large}, etc., for which a very high accuracy and a system size of thousands to hundreds of millions of atoms, or even larger, is often required.
In a longer term, this could be used to problems of significant practical interest, such as drug design and materials design. 
%that are of a more significant practical interest, such as drug design and materials design.
%RC{let's avoid classifying the practical interest: electrochemical cells are also of significant practical interest for example  

\subsection{Outlook in the era of Exascale computing\label{sec:8-2}}
% \LL{ Given Fugaku is out, we might need to beef up this paragraph slightly.}
% \WL{added one paragraph, on the FLOP/Byte ratio. I believe we can have better FLOPS on Fugaku. }
The past decade has witnessed the rapid growth of the many-core architecture due to its superior performance in FLOPS per watt and memory bandwidth. 
This essentially requires a revisit of the scientific applications and a rethinking of the optimal data layout and MPI communication at an algorithmic level,
rather than simply offloading computational intensive tasks.
%using GPUs purely as accelerators by 
In this paper, the critical data layout in DeePMD is redesigned to increase the task granularity, then the entire DeePMD-kit code is parallelized and optimized to improve its scalability and efficiency on the GPU supercomputer Summit. 
The optimization strategy presented in this paper can also be applied to other many-core architectures.
For example, it can be easily converted to the Heterogeneous-compute Interface for Portability (HIP) programming model to run on the next exascale supercomputer Frontier, which will be based on AMD GPUs.

In the pursuit of greater computational power, the computational power v.s.~memory bandwidth ratio (or FLOP/Byte ratio in short) rises rapidly, especially when specialized half-precision hardware is involved.
For example, the double-precision FLOP/Byte ratio on V100 GPU is 7.8, while the half-precision FLOP/Byte ratio is 133.3
%e.g., 
%The FLOP/Byte ratio is 133 for V100 GPU 
($120$TFLOPS/$900$GB/s$=133.3$ FLOP/B), which means the 120 TFLOPS half-precision computing power can only be achieved when 133 operations are executed after a single byte is loaded from global memory into the GPU. 
Such a high ratio makes it difficult to utilize the full computing power of the Tensor Cores with small matrix size, which is exactly in the case of optimized DeePMD-kit --- the MIX-16 version is mainly bounded by the GPU memory bandwidth. 
This implies that future improvement of the FLOP/Byte ratio for the many-core architecture, especially for the half-precision specialized hardware, can benefit HPC+AI applications such as DeePMD-kit. 
We notice that on the newly announced Fugaku supercomputer, the Fujitsu A64FX CPU has a FLOP/Byte ratio of 13.2 ($13.51$TFLOPS/$1024$GB/s$=13.2$FLOP/Byte) in the boost mode. 
Therefore, in theory, the optimized DeePMD-kit should achieve better performance on the Fugaku supercomputer. 
%Given that Fugaku has larger capacity of high bandwidth memory, the computationally feasible system size can also increase accordingly. 
In addition, the computationally feasible system size of the optimized DeePMD-kit can be increased on future many-core architecture if the capacity of the high bandwidth memory is  expanded. 

Based on the scaling shown in Fig.~\ref{fig:weak}, we see no intrinsic obstacles to scaling our code to run on the exascale supercomputer for systems with billions of atoms. 
Compared to the traditional numerical methods such as density functional theory, one advantage of Deep Potential lies in its resilience to numerical noise, which could significantly reduce the amount of work needed for fault-tolerant treatments. 
Therefore, methods like DeePMD can be ideal candidates in the upcoming era of exascale computing.
%\LL{I think we can add a sentence here saying that compared to standard computational models, deep learning models such as DP can be naturally more resilient to numerical noises. This could significantly reduce the amount of work needed for fault-tolerant computing, and hence could be ideal candidates in the coming era of Exascale computing.} \WL{added based on what your wrote.}
%\WH{It is not obvious to me why DP is more resilience to numerical noises than DFT.}\LL{ because there is a training procedure involved, and SGD needs to handle noise anyway? let me know whether this makes sense} \WH{No training on HPC. we perform MD with trained model. the model can be trained by a single GPU card.} \LL{ A crazy idea would be to perform some micro-adjustment of the parameters actually using HPC, or dump out some error model of the actual HPC machine and train it with the error model on a single GPU card to do ``error mitigation''. Basically the largest challenge of Exascale computing is fault-tolerance (other than power usage), so it would be good to say something here} \WL{that is the idea of a fault tolerance paper in SC'18 (if I remember correctly), they train on the log data of actual failures on supercomputers, to predict the next node failure. quite amazing work.}
On the other hand, improvements on the hardware, especially reducing the latency of GPU and network, are required to achieve better strong scaling for the DeePMD-kit on the next generation supercomputers. 
\vskip 1em

\section*{Acknowledgment}
Numerical tests were performed on the Summit supercomputer located in the Oak Ridge National Laboratory. This work was partially supported by the National Science Foundation
under Grant No. 1450372, No. DMS-1652330  (W. J. and L. L.), and by the Department of Energy under Grant No. DE-SC0017867 (L. L.). The work of H. W. is supported by the National Science Foundation of China under Grant No. 11871110, the National Key Research and Development Program of China under Grants No. 2016YFB0201200 and No. 2016YFB0201203, and Beijing Academy of Artificial Intelligence (BAAI). We thank a gift from iFlytek to Princeton University and the ONR grant N00014-13-1-0338 (L. Z. and W. E), and the Center Chemistry in Solution and at Interfaces (CSI) funded by the DOE Award DE-SC0019394 (L. Z., R. C. and W. E). The authors would like to thank Lin-Wang Wang, Chao Yang for helpful discussions, and Junqi Yin, Bronson Messer and the team of ORNL for their support.  

%\section*{References}

\balance

%\bibliography{IEEEabrv, ref}
\bibliography{ref}
\bibliographystyle{IEEEtran}

\end{document}